\documentclass[aps,physrev,reprint,superscriptaddress,nofootinbib]{revtex4-2}

\usepackage{graphicx}
\usepackage{amsmath}
\usepackage{amssymb}
\usepackage{booktabs}
\usepackage{xcolor}
\usepackage{float}
\usepackage{soul}
\usepackage{bm}
\usepackage{colortbl}
\usepackage{url}
\usepackage{algpseudocode}
\usepackage{enumitem}
\usepackage{etoolbox}
\usepackage{makecell}
\usepackage{multirow}
\usepackage{threeparttable}
\usepackage{tablefootnote}

\newcommand{\yes}{\ensuremath{\bullet}}
\newcommand{\no}{\ensuremath{\circ}}

\algrenewcommand\algorithmicrequire{\textbf{Input:}}
\algrenewcommand\algorithmicensure{\textbf{Output:}}
\algrenewcommand{\algorithmiccomment}[1]{\hfill\# #1}

\usepackage{hyperref}
\hypersetup{
  colorlinks=true,
  linkcolor=blue,
  citecolor=blue,
  urlcolor=blue,
}

\begin{document}
\makeatletter
\floatstyle{ruled}
\newfloat{algorithm}{tbp}{loa}
\floatname{algorithm}{Algorithm}
\makeatother

\makeatletter
\renewcommand\fs@ruled{%
  \def\@fs@cfont{}%
  \let\@fs@capt\floatc@ruled
  \def\@fs@pre{}%
  \def\@fs@mid{\kern2pt\hrule height0.8pt\kern2pt\hrule height0.8pt\kern2pt}%
  \def\@fs@post{\kern2pt\hrule height0.8pt\kern2pt\hrule height0.8pt\relax}%
  \let\@fs@iftopcapt\iftrue}
\renewcommand{\floatc@ruled}[2]{{\@fs@cfont \raggedright #1.\hskip\labelsep #2\par}}
\renewcommand{\fnum@algorithm}{\fname@algorithm~\thealgorithm}
\makeatother

\title{KIGNet: Physics-Motivated Multi-Graph Representation Learning for Explainable Jet Tagging}

\author{Md Raqibul Islam}
\thanks{These two authors contributed equally to this work}
\email[Contact author: ]{raqibul.islam.academic@gmail.com}
\affiliation{Center for Computational \& Data Sciences, Independent University, Bangladesh, Dhaka-1229, Bangladesh}

\author{Adrita Khan}
\thanks{These two authors contributed equally to this work}
\affiliation{Center for Computational \& Data Sciences, Independent University, Bangladesh, Dhaka-1229, Bangladesh}

\author{Mir Sazzat Hossain}
\email[Contact author: ]{sazzat@iub.edu.bd}
\affiliation{Center for Computational \& Data Sciences, Independent University, Bangladesh, Dhaka-1229, Bangladesh}

\author{Choudhury Ben Yamin Siddiqui}
\affiliation{Center for Computational \& Data Sciences, Independent University, Bangladesh, Dhaka-1229, Bangladesh}

\author{Md.\ Zakir Hossan}
\affiliation{Center for Computational \& Data Sciences, Independent University, Bangladesh, Dhaka-1229, Bangladesh}

\author{Tanjib Khan}
\affiliation{Department of Theoretical Physics, University of Dhaka, Dhaka-1000, Bangladesh}

\author{M.\ Arshad Momen}
\affiliation{Center for Computational \& Data Sciences, Independent University, Bangladesh, Dhaka-1229, Bangladesh}
\affiliation{Department of Physical Sciences, Independent University, Bangladesh, Dhaka-1229, Bangladesh}

\author{Amin Ahsan Ali}
\affiliation{Center for Computational \& Data Sciences, Independent University, Bangladesh, Dhaka-1229, Bangladesh}

\author{AKM Mahbubur Rahman}
\affiliation{Center for Computational \& Data Sciences, Independent University, Bangladesh, Dhaka-1229, Bangladesh}

\begin{abstract}
Jet identification and classification play a central role in the analysis of data from high-energy collider experiments. While deep learning has improved jet classification, it often lacks interpretability. We introduce the Kinematic Interaction Graph Network (KIGNet), a graph neural network that integrates kinematic variables into jet classification by constructing four graph representations per jet, each weighted by a distinct variable: angular separation ($\Delta$), relative transverse momentum ($k_T$), momentum fraction ($z$), and invariant mass squared ($m^2$). Three of these variables ($\Delta$, $k_T$, $z$) are motivated by the Lund jet plane, a framework grounded in perturbative QCD factorization; the fourth ($m^2$) provides complementary mass-scale sensitivity for heavy-flavor jet identification. Using the concept of Gradient-weighted Class Activation Mapping (Grad-CAM), we determine which kinematic variables dominate classification outcomes. Angular separation and relative transverse momentum collectively account for approximately 76\% of the total Grad-CAM attribution (40.72\% and 35.67\%, respectively), with momentum fraction and invariant mass contributing the remaining 24\%. This hierarchy is consistent with the soft-collinear structure of QCD radiation encoded in the training data, demonstrating that the network learns physically interpretable representations rather than spurious correlations. Evaluated on the JetClass dataset, KIGNet achieves a macro-accuracy of 95.07\%, macro-AUC of 96.61\%, and macro-AUPR of 81.52\%, representing relative improvements of 2.45\%, 3.40\%, and 19.11\%, respectively, over the state-of-the-art baseline model. Evaluated on the Aspen Open Jets dataset of real CMS collision data, KIGNet produces substantially more structured latent representations than the baseline, reducing the Davies-Bouldin Index by 52.15\% ($0.8395 \rightarrow 0.4017$) and increasing the Dunn Index by 42.33\% ($0.0189 \rightarrow 0.0269$), confirming that physics-informed kinematic encoding generalizes beyond idealized simulation to experimental detector conditions.
\end{abstract}

\maketitle

\section{Introduction}

\label{sec:intro}

At the Large Hadron Collider (LHC), most high-energy collisions produce jets, collimated sprays of hadrons formed as energetic quarks and gluons fragment and hadronize. As short-distance partons are not directly observable, the properties of high-energy collisions must be reconstructed from the resulting jets. Identifying the particle that initiated a jet, known as jet tagging, is therefore important for collider physics analyses at the Large Hadron Collider~\cite{Larkoski2018}. Such discrimination is essential for precision Standard Model measurements, from distinguishing light-quark and gluon jets to identifying boosted hadronic decays of Higgs bosons, top quarks, and electroweak gauge bosons~\cite{Mondal2024_ML4HEP, Albertsson2018_ML4HEP}. The challenge will only grow: the High-Luminosity LHC (HL-LHC) is expected to raise the integrated luminosity by about an order of magnitude relative to the current program~\cite{aberle2020, CERN:HLLHC}, so taggers must extract more information from each event while operating under increasingly tight computational and latency budgets~\cite{Qu2020_ParticleNet, Qu2022_ParticleTransformer}.

Jet tagging is not a single problem but a collection of related classification tasks, each tied to distinct physics objectives. Light-quark versus gluon discrimination separates the two most abundant jet types, which differ in their radiation profiles due to their differing color charges. Heavy-flavor tagging identifies $b$- and $c$-initiated jets through displaced secondary vertices and the suppression of small-angle radiation. When a heavy resonance is produced at large transverse momentum, its decay products merge into a single large-radius jet; in this regime, the identification of $W$-boson, $Z$-boson, Higgs-boson, and top-quark jets reduces to resolving multi-prong substructure~\cite{Larkoski2018}. What unites these tasks is that the discriminating information lives in the pattern of QCD radiation inside the jet, governed by well-understood splitting dynamics yet difficult to capture with any single handcrafted variable.

\begin{figure*}[htbp]
    \centering
    \includegraphics[width=\textwidth,keepaspectratio]{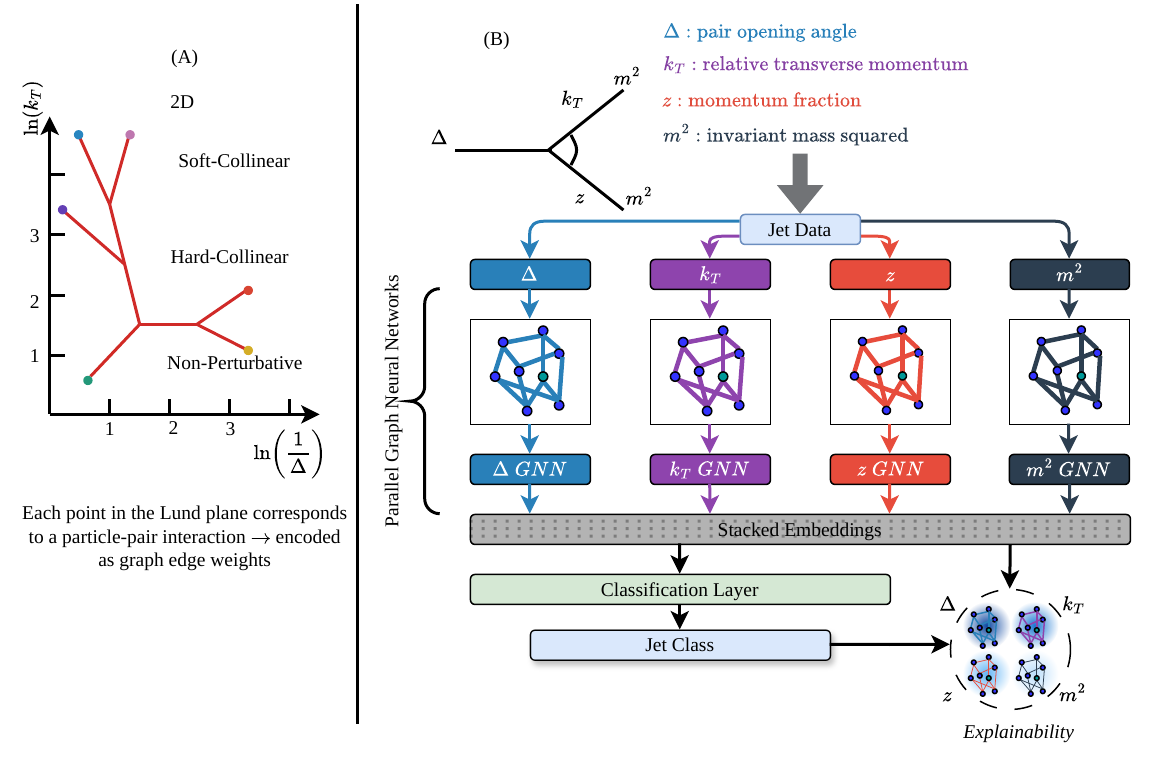}
    \caption{(A) Each particle-pair interaction is mapped onto the Lund plane, where soft-collinear, hard-collinear, and non-perturbative regions encode the underlying splitting kinematics and serve as graph edge weights. The plane follows the convention of~\cite{Dreyer2018_LundJetPlane}; KIGNet uses $\ln\Delta$ rather than $\ln(1/\Delta)$ (a sign flip with no loss of expressive power; see Section~\ref{subsec:kinematic-mg}). (B) Four kinematic variables, angular separation ($\Delta$), relative transverse momentum ($k_T$), momentum fraction ($z$), and invariant mass squared ($m^2$) (Eqs.~\eqref{eq:delta_r}-\eqref{eq:invariant_mass}), define parallel edge-weighted graph views processed by four independent GNNs. The resulting embeddings are stacked and passed to a classification layer, with an auxiliary branch providing Grad-CAM explainability (Section~\ref{sec:gradcam}). Full architecture details are given in Figure~\ref{fig:model_architecture} and Section~\ref{subsec:network_architecture}.}
    \label{fig:overview}
\end{figure*}

Progress in jet tagging with machine learning has followed a sequence of increasingly detailed jet representations, each designed to overcome limitations of its predecessor. The earliest taggers reduced a jet to a few handcrafted substructure observables, such as $N$-subjettiness~\cite{Thaler2011_Nsubjettiness} and the energy correlation functions~\cite{Larkoski2013EnergyCorrelation}, which are infrared- and collinear-safe and physically transparent but compress the full constituent information into a handful of scalars at the cost of discriminating power. Deep learning first accessed lower-level information by treating the jet as a fixed-resolution calorimeter image processed by convolutional networks~\cite{Cogan2015_JetImages,Oliveira2015_jetimages}, though pixelization and the sparsity of energy deposits limited performance and did not naturally accommodate the variable particle multiplicity of a jet. Sequence-based models followed, with recursive networks built on the jet clustering tree~\cite{Louppe2017_RecNN} and a one-dimensional convolutional tagger applied to $p_T$-ordered constituents~\cite{cms2017boosted}, but both imposed an artificial ordering on what is intrinsically an unordered set. Deep Sets~\cite{Zaheer2017_DeepSets} removed this ordering by aggregating per-particle features symmetrically, and the Energy and Particle Flow Networks brought the same permutation-invariant principle to jets~\cite{Komiske2019_EFN}, at the cost of not modeling explicit pairwise correlations between particles. Graph- and point-cloud-based networks reinstated those relations: ParticleNet learns on a dynamically constructed particle cloud with edge convolutions~\cite{Qu2020_ParticleNet}, JEDI-net models explicit pairwise interactions through an interaction network~\cite{Moreno2019_JEDINet}, and LundNet builds its graph directly from the Lund declustering tree~\cite{Dreyer2021_Lund_Plane}. Related graph and point-cloud taggers include particle convolutions~\cite{Shimmin2021_ParticleConv}, point-cloud transformers~\cite{Mikuni2021_PointCloud}, graph networks with Haar pooling~\cite{Ma2022_HaarPooling_GN}, and graph networks for boosted Higgs jet reconstruction~\cite{Guo_Boosted_Higss_GNN}. A parallel line of work encodes physical priors and exact symmetries into the architecture, beginning with the Lund jet plane as a QCD-motivated coordinate system~\cite{Dreyer2018_LundJetPlane} and continuing with Lorentz-equivariant networks such as LGN~\cite{Bogatskiy2020_LGN}, LorentzNet~\cite{Gong2022_Lorentz_GNN}, and PELICAN~\cite{Bogatskiy2022_PELICAN}, which match or exceed earlier accuracy with far fewer parameters by respecting the symmetries of the underlying physics. Attention-based transformers now define the state of the art, with the Particle Transformer incorporating learned pairwise interaction features into self-attention; the accompanying JetClass benchmark~\cite{Qu2022_ParticleTransformer,Qu2022_JetClassDataset} has since become a standard point of comparison. Most recently, the latency constraints of the HL-LHC trigger have motivated a parallel line of work prioritizing inference speed, including the Particle Chebyshev Network (PCN), which approximates graph convolutions with Chebyshev polynomials~\cite{Semlani2024PCN}, the sub-microsecond JEDI-linear~\cite{Que2025_JEDI_linear}, and the linear-attention SAL-T~\cite{Wang2025_SAL_T}, each accepting a small accuracy trade-off in exchange for real-time deployability.

Despite this progress, the most accurate models behave as black boxes, making it difficult to determine whether they exploit genuine QCD radiation structure or correlations specific to their training samples. This concern is sharpened for spectral methods such as PCN, our strongest baseline, where it is difficult to trace how the graph structure shapes the learned representation~\cite{Wetzel2025}. Compounding this, evaluation is almost always confined to simulation, leaving robustness on real detector data, with its pileup and reconstruction effects, largely untested. As machine-learning taggers move from benchmark studies toward precision measurements and new-physics searches, representations grounded in established QCD principles that generalize beyond simulation become essential.

To address these limitations, we introduce the Kinematic Interaction Graph Network (KIGNet), a graph neural network that augments the jet graph with explicit pairwise kinematic interactions grounded in QCD. For every pair of connected particles, KIGNet computes four kinematic variables\footnote{Throughout this paper, we use the terms ``interaction features'', ``kinematic variables'', and ``kinematic edge features'' interchangeably.}: angular separation ($\Delta$), relative transverse momentum ($k_T$), momentum fraction ($z$), and invariant mass squared ($m^2$). The first three are motivated by the Lund jet plane and perturbative QCD factorization~\cite{Dreyer2018_LundJetPlane, Dreyer2021_Lund_Plane}, while $m^2$ adds mass-scale sensitivity for heavy-flavor jets. As illustrated in Figure~\ref{fig:overview}, each variable defines a separate edge-weighted view of the jet, the four views are processed by parallel graph neural network branches, and their embeddings are combined for classification, so that the network learns a specialized representation for each physical aspect of jet formation rather than forcing one network to balance all of them at once. To probe model interpretability, we adapt Gradient-weighted Class Activation Mapping (Grad-CAM)~\cite{Selvaraju2017_GradCAM} to this multi-graph setting, obtaining a quantitative measure of how strongly each kinematic variable drives the classification decision.

We evaluate KIGNet on the JetClass benchmark, taking the Particle Chebyshev Network as our primary baseline and comparing against PFN~\cite{Komiske2019_EFN}, P-CNN~\cite{cms2017boosted}, ParticleNet~\cite{Qu2020_ParticleNet}, and ParT~\cite{Qu2022_ParticleTransformer}. KIGNet achieves a macro-averaged accuracy of $95.07\%$, a macro-AUC of $96.61\%$, and a macro-AUPR of $81.52\%$, improving over the PCN baseline by $2.45\%$, $3.40\%$, and $19.11\%$, respectively, and surpassing it across all signal classes. The Grad-CAM analysis reveals a physically interpretable hierarchy in which angular separation and relative transverse momentum together account for approximately $76\%$ of the total attribution score, consistent with the dominant role of soft-collinear radiation in QCD, while the smaller contributions of $z$ and $m^{2}$ reflect energy-sharing dynamics and heavy-quark mass effects. To test generalization beyond simulation, we evaluate on the Aspen Open Jets dataset of real CMS proton-proton collision data~\cite{Amram2024AOJ}, where KIGNet reduces the Davies-Bouldin Index by $52.15\%$ and increases the Dunn Index by $42.33\%$ relative to the baseline, indicating substantially more structured representations under realistic detector conditions.

\textbf{Summary of contributions.}
\begin{itemize}
    \item \textbf{Physics-motivated multi-graph architecture.} We propose KIGNet, which constructs four parallel, edge-weighted graph views of each jet, weighted by the kinematic variables $\Delta$, $k_T$, $z$, and $m^2$, so that the network can specialize to complementary aspects of the jet's radiation pattern.

    \item \textbf{Grad-CAM explainability for multi-graph GNNs.} We adapt Grad-CAM to the multi-graph setting and obtain a quantitative importance ranking of the kinematic variables, finding that $\Delta$ and $k_{T}$ together receive approximately $76\%$ of the total Grad-CAM attribution, qualitatively consistent with expectations from soft-collinear QCD radiation.

    \item \textbf{State-of-the-art classification performance.} KIGNet achieves a macro accuracy of 95.07\%, macro AUC of 96.61\%, and macro AUPR of 81.52\% on the JetClass benchmark, outperforming a strong PCN baseline across all signal classes in accuracy.

    \item \textbf{Generalization to real collider data.} On the Aspen Open Jets dataset of real CMS collisions~\cite{Amram2024AOJ}, KIGNet reduces the Davies-Bouldin Index by 52.15\% and increases the Dunn Index by 42.33\% relative to the baseline, demonstrating more structured representations under realistic detector effects, pileup, and reconstruction uncertainties.
\end{itemize}

The remainder of this paper is organized as follows. Section~\ref{sec:related} reviews related work on jet tagging together with the relevant QCD and Lund-plane background. Section~\ref{sec:method} presents the KIGNet framework, including the dataset, graph representation, network architecture, and training configuration. Section~\ref{sec:results} reports the classification results, model comparisons, and Grad-CAM analysis. Section~\ref{sec:ablation} presents ablation studies, followed by discussion in Section~\ref{sec:dis} and conclusions in Section~\ref{sec:conclusion}.

\section{Physics Background and Previous Work}
\label{sec:related}

The architectural evolution of machine-learning jet taggers, from calorimeter images and ordered particle sequences, through permutation-invariant set and graph representations, to attention-based and symmetry-respecting networks, is reviewed in Section~\ref{sec:intro}. Here we focus on the QCD foundations that motivate KIGNet's kinematic graph representations, beginning with how pairwise interaction variables have entered recent tagger designs before turning to the Lund-plane formalism underlying our approach.

Recent architectures have also begun treating pairwise kinematic quantities as learnable interaction embeddings, passed through dedicated neural modules and combined at the jet level, rather than being used as fixed handcrafted inputs~\cite{Qu2022_ParticleTransformer, Dreyer2021_Lund_Plane, Dreyer2018_LundJetPlane}. Unlike prior works that focus primarily on simulation benchmarks, we explicitly evaluate representation quality on real collider data to assess robustness under domain shift.

\subsection{Jet Tagging and QCD}

Jets originate from QCD parton branching and hadronization, and differences between quark-, gluon-, heavy-flavor-, and boosted heavy-particle-initiated jets are encoded in their radiation patterns. Jet-tagging methods therefore seek observables that are sensitive to the underlying QCD splitting dynamics while remaining robust to detector effects and non-perturbative corrections~\cite{Larkoski2018}.

QCD provides the theoretical framework for understanding jet formation, evolution, and substructure through perturbative calculations, resummation techniques, and parton shower models~\cite{Ali2010, Larkoski2018}. Jets are collimated bundles of hadrons that reflect quark and gluon configurations at short distances, and their observation played a crucial role in establishing QCD as the theory of strong interactions~\cite{ellis2010jets}.

Jet substructure observables are grounded in QCD resummation, which predicts energy profiles and mass distributions by resumming large logarithmic contributions to all orders in the coupling constant~\cite{Li:2011hy, Larkoski2013EnergyCorrelation}. Experimentally, jet tagging leverages both QCD-inspired observables such as $N$-subjettiness, jet angularities, and the Lund plane~\cite{Bhattacherjee:2022gjq, Bertolini:2017efs} and data-driven machine learning approaches to discriminate signal from background. Modern deep learning architectures, particularly GNNs like ParticleNet~\cite{Qu2020_ParticleNet} and transformer-based models like ParT~\cite{Qu2022_ParticleTransformer}, achieve state-of-the-art performance by learning QCD radiation patterns, subjet structures, and particle correlations. Heavy-flavor jet tagging is especially important for testing perturbative QCD, as the large $b$-quark mass suppresses soft-gluon radiation at small angles~\cite{Dokshitzer1991DeadCone}. Interpretability studies reveal that successful jet taggers learn traditional QCD substructure features such as jet multiplicity, width, and mass distributions, with different classification tasks exhibiting distinct sensitivity patterns reflecting underlying QCD radiation mechanisms~\cite{Moreno2019_JEDINet}.

Among the many QCD-based descriptions of jet radiation, the Lund jet plane provides a particularly useful representation because it organizes emissions according to their angular and transverse-momentum scales. The variables entering the Lund plane arise directly from the factorized QCD emission probability and therefore furnish physically interpretable descriptors of jet evolution. Since KIGNet is built upon these quantities, we briefly review the Lund-plane formalism below.

\subsection{Lund Plane Formalism}

The kinematic variables used in this work are motivated by the Lund jet plane~\cite{Dreyer2018_LundJetPlane}, which provides a two-dimensional representation of jet substructure by mapping iterative declustering onto the $(\ln k_T, \ln \Delta)$ plane.

In perturbative QCD, the differential emission probability factorizes as
\begin{equation}
\label{eq:emission_full}
dP \propto \alpha_s(k_T)\, \frac{dk_T}{k_T}\, \frac{d\Delta}{\Delta}\, P(z)\, dz,
\end{equation}
where $P(z)$ denotes the leading-order DGLAP splitting function (see Appendix~\ref{app:qcd_theory} for explicit expressions). In the soft limit $z \to 0$, the splitting functions behave as $P(z) \to C_R/z$, where $C_R$ is the relevant color factor ($C_F = 4/3$ for quarks, $C_A = 3$ for gluons). Substituting $P(z) \to C_R/z$ into Eq.~\eqref{eq:emission_full} and integrating over $z \in (0, z_{\mathrm{cut}})$ introduces only a logarithmic prefactor that is constant across the Lund plane, yielding the two-dimensional soft-limit form
\begin{equation}
\label{eq:emission_soft}
dP \propto \alpha_s\, d(\ln k_T)\, d(\ln \Delta),
\end{equation}
which is equivalent to the symmetric logarithmic form $dP \propto \alpha_s\, d(\ln k_T)\, d(\ln \Delta)\, d(\ln z)$ familiar from the original Lund plane construction~\cite{Dreyer2018_LundJetPlane} when the full $z$ dependence is retained. This factorization makes the Lund plane a natural coordinate system for jet radiation patterns and an effective framework for capturing differences between quark-, gluon-, and boosted heavy-particle-initiated jets.

The Lund plane is constructed via iterative declustering of a jet reclustered using the Cambridge/Aachen (C/A) algorithm~\cite{Dokshitzer:1997in, Wobisch:1998wt}, which clusters particles by angular separation and thereby encodes a geometrically ordered emission history. Beginning from the full jet $j$, the last clustering step is undone to obtain two subjets $j_1$ and $j_2$ with $p_{T,1} > p_{T,2}$. 

Each splitting $i \to j + k$ is described by four variables derived from 
particle four-momenta $p = (E, p_x, p_y, p_z)$:
\begin{align}
\Delta_{ab} &= \sqrt{(y_a - y_b)^2 + (\Delta\phi_{ab})^2} 
\label{eq:delta_r} \\
k_{T,ab} &= \min(p_{T,a}, p_{T,b}) \cdot \Delta_{ab} 
\label{eq:k_t} \\
z_{ab} &= \frac{\min(p_{T,a}, p_{T,b})}{p_{T,a} + p_{T,b}} 
\label{eq:momentum_fraction} \\
m_{ab}^2 &= (E_a + E_b)^2 - \lVert \vec{p}_a + \vec{p}_b \rVert^2 
\label{eq:invariant_mass}
\end{align}
where
\[
y_i = \frac{1}{2}\ln\frac{E_i + p_{z,i}}{E_i - p_{z,i}}
\]
is the rapidity, $\phi_i$ is the azimuthal angle,
\[
\Delta\phi_{ab} = \bigl[(\phi_a - \phi_b + \pi)\bmod 2\pi\bigr] - \pi
\]
is the azimuthal angle difference wrapped to the interval $[-\pi, \pi]$,
\[
p_{T,i} = \sqrt{p_{x,i}^2 + p_{y,i}^2}
\]
is the transverse momentum, and $\vec{p}_i = (p_{x,i}, p_{y,i}, p_{z,i})$ 
is the three-momentum. The quantity 
$m_{ab}^2 = (E_a + E_b)^2 - \lVert \vec{p}_a + \vec{p}_b \rVert^2$ 
is the Lorentz-invariant mass squared of the particle pair $(a,b)$, 
computed directly from their four-momenta.

The procedure continues by setting $j \to j_1$ and repeating until only a single particle remains, tracing the hardest branch of the jet's splitting history. Each step is mapped onto a point in the Lund plane spanned by $\ln k_T$ and $\ln(1/\Delta)$, following the convention of~\cite{Dreyer2018_LundJetPlane}. In our implementation we use $\ln\Delta$ rather than $\ln(1/\Delta)$ as the angular edge feature (a sign flip), which is equivalent in expressive power since the GNN applies learned affine transformations to its inputs.

These variables exhibit long-tailed distributions spanning multiple orders of magnitude. Following Dreyer \& Qu~\cite{Dreyer2021_Lund_Plane}, we use the kinematic variables' logarithmic forms ($\ln \Delta$, $\ln k_T$, $\ln z$, $\ln m^2$) as interaction features to improve numerical stability and distribution normality during training, consistent with the logarithmic measure $d(\ln k_T)\, d(\ln \Delta)\, d(\ln z)$ arising from the full emission probability~\eqref{eq:emission_full} in the soft limit~\eqref{eq:emission_soft}.

The physical significance of each variable reflects distinct QCD dynamics: $\Delta$ encodes angular ordering and collinear emissions; $k_T$ sets the relative transverse momentum scale determining $\alpha_s(k_T)$ and separating perturbative from non-perturbative regimes; $z$ quantifies energy sharing between daughter partons via the DGLAP splitting functions $P(z)$~\cite{Dokshitzer1977DGLAP, GRIBOV1971_DGLAP}; and $m^2$ provides mass-scale sensitivity essential for identifying heavy-flavor jets where quark masses become non-negligible. Among these, $\Delta$ and $k_T$ are the natural axes of the standard Lund plane~\cite{Dreyer2018_LundJetPlane}; $z$ is not a Lund plane axis but appears in the emission probability~\eqref{eq:emission_full} and is used as an additional Lund motivated feature following~\cite{Dreyer2021_Lund_Plane}; and $m^2$ is an additional variable not part of the Lund plane formalism but included for its sensitivity to heavy quark mass effects.

To leading order in perturbative QCD, soft-collinear emissions are distributed uniformly in the Lund plane with density $\rho \propto \alpha_s\, \frac{dz}{z}\, \frac{d\theta}{\theta}$, a consequence of angular ordering and soft-collinear factorization. This structure permits a clean separation of the dynamical regimes of jet formation: the hard-collinear region (large $z$) captures dominant jet splitting; the soft-collinear region corresponds to small $z$ and small $\Delta$; initial-state radiation manifests at large angular separations; and the non-perturbative regime is confined to $k_T \lesssim 0.5\,\text{GeV}$. The Lund plane has proven effective for a broad class of jet-physics applications, subsumes groomed jet observables including those from soft-drop declustering into a unified framework, and serves as a natural input space for machine-learning-based taggers.

In KIGNet, we incorporate these kinematic variables through a multi-graph architecture. The four variables $\Delta$, $k_T$, $z$, and $m^2$ encode complementary aspects of QCD dynamics: geometric structure, radiative scales, splitting probabilities, and mass thresholds, making them natural candidates for multi-graph representation learning.

\section{Methodology}
\label{sec:method}

This section describes KIGNet, our proposed approach for jet classification. We introduce the datasets in Section~\ref{subsec:dataset}, explain the graph representation in Section~\ref{subsec:jets_graph_representation}, describe the physics-motivated edge-weighted multi-graph architecture in Section~\ref{subsec:kinematic-mg}, present the neural network architecture in Section~\ref{subsec:network_architecture}, and conclude with training configuration in Section~\ref{subsec:training_config}.

\subsection{Datasets}
\label{subsec:dataset}

\subsubsection{JetClass}
\label{subsubsec:JetClass}

We conducted our experiments on the \texttt{JetClass} dataset~\cite{Qu2022_ParticleTransformer, Qu2022_JetClassDataset}, a large-scale benchmark designed to advance deep learning research in jet physics. \texttt{JetClass} comprises 100 million jets for training, 5 million for validation, and 20 million for testing, distributed across 10 jet classes: 9 signal classes and 1 background. The 9 signal classes comprise Higgs boson decays ($H \to b\bar{b}$, $H \to c\bar{c}$, $H \to gg$, $H \to 4q$, $H \to \ell\nu qq'$), top quark decays ($t \to bqq'$, $t \to b\ell\nu$), and electroweak boson decays ($W \to qq'$, $Z \to q\bar{q}$); the background class consists of jets from light quarks and gluons (q/g). Table~\ref{tab:signal_classes} presents the complete taxonomy.

\begin{table}[htbp]
\centering
\caption{The 10 jet classes in the JetClass dataset~\cite{Qu2022_ParticleTransformer}. Classes 1-9 are signal jets arising from heavy particles (Higgs, $W$, $Z$ bosons, and top quarks), while class 10 ($q/g$) corresponds to background jets initiated by light quarks or gluons.}
\label{tab:signal_classes}
\begin{tabular*}{\columnwidth}{@{\extracolsep{\fill}}p{1cm} p{1.7cm} p{5.0cm}}
\toprule \toprule
Class & Decay Process & Description \\
\midrule
1  & $H \rightarrow b\bar{b}$   & Higgs boson decays to bottom quark-antiquark pair \\
2  & $H \rightarrow c\bar{c}$   & Higgs boson decays to charm quark-antiquark pair \\
3  & $H \rightarrow gg$         & Higgs boson decays to two gluons \\
4  & $H \rightarrow 4q$         & Higgs boson decays to four quarks \\
5  & $H \rightarrow \ell\nu qq'$   & Higgs boson decays to lepton, neutrino, and two quarks \\
6  & $t \rightarrow bqq'$       & Top quark decays to bottom quark and two other quarks \\
7  & $t \rightarrow b\ell\nu$      & Top quark decays to bottom quark, lepton, and neutrino \\
8  & $W \rightarrow qq'$        & $W$ boson decays to two quarks \\
9  & $Z \rightarrow q\bar{q}$   & $Z$ boson decays to quark-antiquark pair \\
10 & $q/g$                       & Background jets initiated by light quarks or gluons \\
\bottomrule \bottomrule
\end{tabular*}
\end{table}

For computational efficiency while preserving class balance, we use a subset of 1 million training jets (100,000 per class). The 1M jet events were divided into 800k for training, 100k for validation, and 100k for an initial test set. The final performance metrics are computed using the complete test set of 20M jets. Table~\ref{tab:data_split} details the dataset partitioning.

\begin{table}[htbp]
\centering
\caption{Data used for model development and evaluation.}
\label{tab:data_split}
\begin{tabular*}{\columnwidth}{@{\extracolsep{\fill}}lrr}
\toprule \toprule
Subset & Total Jets & Per Class \\
\midrule
Training & 1,000,000 & 100,000 \\
Test & 20,000,000 & 2,000,000 \\
\bottomrule \bottomrule
\end{tabular*}
\end{table}

\paragraph{Treatment of the Background Class.}
The background class (q/g) is included during training and validation. The output layer produces ten logits and the cross-entropy loss is computed over all ten classes. So, q/g jets constitute a genuine tenth category that the network must actively discriminate from each signal class.

Per-class results (Table~\ref{tab:jetclass_performance}) and all macro-averaged metrics are reported over the signal classes only, consistent with the JetClass benchmark convention~\cite{Qu2020_ParticleNet, Qu2022_ParticleTransformer}. Including the background in macro-averaging would inflate performance, since q/g jets are typically easier to reject than to distinguish among signal classes; this convention is applied uniformly across all compared models. Note that per-class AUC and AUPR values nonetheless encode background rejection implicitly: under one-vs.-rest binarization the negative class pools all remaining classes, including q/g, so the strong AUC values in Table~\ref{tab:jetclass_performance} reflect both inter-signal discrimination and background rejection simultaneously.

\subsubsection{Aspen Open Jets}
\label{subsubsec:AOJ}

Aspen Open Jets (AOJ)~\cite{Amram2024AOJ} is a large-scale, machine-learning-ready dataset of hadronic jets constructed from the CMS 2016 JetHT proton-proton collision Open Data at the LHC. It comprises approximately $170$-$180$ million high-$p_T$ jets (predominantly QCD, with an expected $\mathcal{O}(10^5)$ $W$, $Z$, and top jets) and provides, for each jet, low level reconstructed particle information together with high level kinematic quantities such as jet transverse momentum ($p_T$), rapidity ($y$), and azimuth ($\phi$). The dataset is built by selecting inclusive jet triggers from the CMS 2016 Open Data and applying quality, kinematic, and detector level selections to ensure stable, well reconstructed jets while preserving the natural mixture of Standard Model processes, thereby capturing realistic detector effects, underlying event, and pileup conditions. The dataset spans five physics categories: QCD, $W$, $Z$, top, and $H\to b\bar{b}$ jets as estimated from CMS Monte Carlo simulations~\cite{Amram2024AOJ}.

For our analysis, we randomly select a subset of 2 million jets from the full dataset and partition it into training, validation, and test sets using an 80:10:10 split, corresponding to 1.6 million, 0.2 million, and 0.2 million jets, respectively.

\subsection{Graph Representation of Jets}
\label{subsec:jets_graph_representation}

We represent each jet as a graph $G = (V, E)$ where particles are nodes $V$ and connections among nearby particles form edges $E$. Each particle is connected to its $k=3$ nearest neighbors based on angular separation in $(\eta, \phi)$ space, where $\eta$ is pseudorapidity and $\phi$ is the azimuthal angle.

\textbf{Choice of $k$ for $k$-Nearest Neighbors.} Excessively high values of $k$ result in denser graphs with more edges per node, fostering increased interconnectivity among particles and potentially introducing noise from distant, weakly related particles. Conversely, excessively low values of $k$ yield sparser graphs that may miss important local interactions. Moreover, $k$ must be strictly less than the number of particles in a jet; higher values of $k$ therefore exclude smaller jets from training altogether. Notably, when $k$ is set within the range of $7$-$10$ (the value suggested by the elbow method), a proportion of the $H \to \ell \nu qq'$ jets is omitted from the training dataset. Preserving these smaller jets ensures the model's proficiency in discerning subtle interactions within collisions. Using the elbow method while avoiding the omission of jets, we find the optimal value $k = 3$.

The nearest neighbor search uses a KD-tree for computational efficiency~\cite{maneewongvatana1999analysis}, which creates a binary tree for quick lookup of nearest neighbors using a dimensional heuristic. Each node's three nearest neighbors are retrieved and linked with an edge, and the completed graph is output as an adjacency matrix for use in training.

Each particle node carries 16 features: momentum components $(p_x, p_y, p_z)$, energy $E$, transverse momentum $p_T$, angular coordinates $(\eta, \phi)$, impact parameters (measuring displacement from the collision point), and one-hot encoded flags indicating particle type (charged hadron, neutral hadron, photon, electron, or muon). These features are concatenated into a single 16-dimensional vector per particle. Table~\ref{tab:input_features} lists all input features.

\subsection{Edge-Weighted Multi-Graph}
\label{subsec:kinematic-mg}

Inspired by recent jet tagging architectures that treat pairwise kinematic quantities as learnable interaction embeddings rather than fixed inputs, we encode the four Lund plane motivated variables $\Delta$, $k_T$, $z$, and $m^2$ as separate graph channels~\cite{Qu2022_JetClassDataset}. Each graph branch processes these interaction features through its own GNN, producing a 64-dimensional jet-level embedding, and the four resulting embeddings are stacked and combined by a $1\times 1$ convolution to form the final representation used for classification. Conceptually, this approach constructs four interaction graphs and learns a separate embedding for each before combining them at the jet level, rather than concatenating $\Delta$, $k_T$, $z$, and $m^2$ directly to node features.

The four kinematic variables are computed for each pair of connected particles according to Equations~\eqref{eq:delta_r}-\eqref{eq:invariant_mass}. We take their logarithms and use $(\ln \Delta,\, \ln k_T,\, \ln z,\, \ln m^2)$ as the kinematic edge features, since these quantities vary over many orders of magnitude and their log-transformed versions exhibit better-behaved distributions during training. Note that the standard Lund plane uses $\ln(1/\Delta)$ on the angular axis~\cite{Dreyer2018_LundJetPlane}; we use $\ln\Delta$ instead (a sign flip) for simplicity of implementation, which does not affect expressive power since the GNN learns its own affine transformation of the input features.

We create four separate graph representations of the same jet using these kinematic variables. All four graphs share identical node features (the 16 particle properties) and the same connectivity (edges to nearest neighbors), but each graph uses a different quantity as its edge weight: $G_{\Delta}$ weighted by $\ln \Delta$, $G_{k_T}$ weighted by $\ln k_T$, $G_{z}$ weighted by $\ln z$, and $G_{m^2}$ weighted by $\ln m^2$. This design allows the network to learn specialized representations for each type of particle relationship rather than requiring a single network to balance all four aspects simultaneously (Algorithm~\ref{alg:multigraph_construction}).

\begin{algorithm}[htbp]
\caption{Multi-Graph Construction from a Jet}
\label{alg:multigraph_construction}
\begin{algorithmic}[1]
\Require Jet of $n$ particles, each with 16 features (Table~\ref{tab:input_features})
\Ensure Weighted graphs $G_{\Delta}, G_{k_T}, G_z, G_{m^2}$
\State $G \gets k$-NN graph ($k{=}3$) on $(\eta,\phi)$ via KD-tree
\For{each edge $(a,b)$ of $G$}
    \State Compute $\ln\Delta,\ \ln k_T,\ \ln z,\ \ln m^2$ from the four-momenta (Eqs.~\eqref{eq:delta_r}-\eqref{eq:invariant_mass})
\EndFor
\For{each $w \in \{\Delta, k_T, z, m^2\}$}
    \State $G_w \gets G$ with edge weights set to $\ln w$
\EndFor
\State \Return $G_{\Delta}, G_{k_T}, G_z, G_{m^2}$
\end{algorithmic}
\end{algorithm}

\subsection{Neural Network Architecture}
\label{subsec:network_architecture}

\begin{figure*}[htbp]
    \centering
    \includegraphics[width=\textwidth]{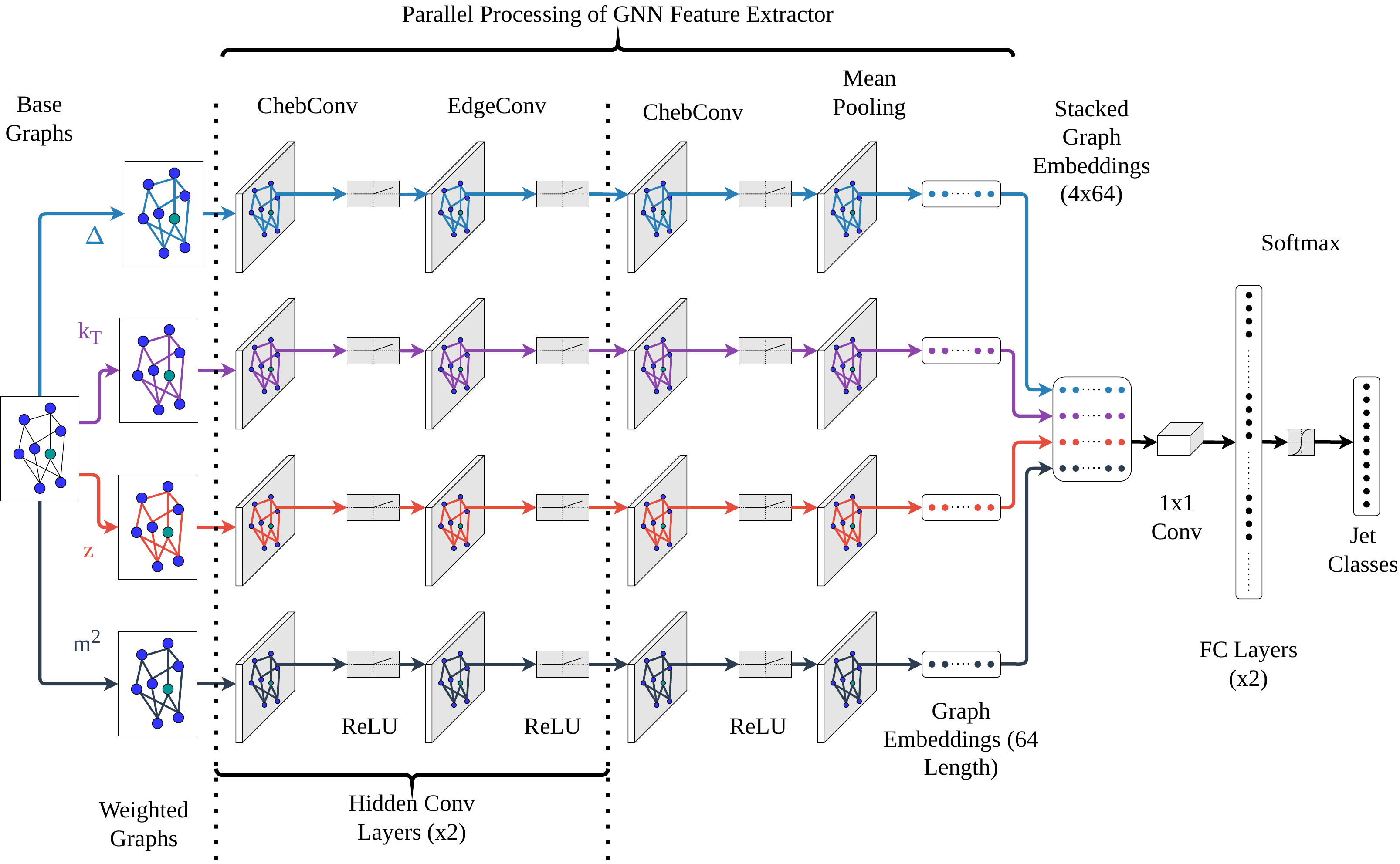}
    \caption{Architecture of the proposed KIGNet model showing four parallel graph processing branches ($G_{\Delta}$, $G_{k_T}$, $G_{z}$, $G_{m^2}$) with hybrid convolutional layers. Each branch processes a distinct graph representation through alternating Chebyshev graph convolutions (ChebConv) and edge convolutions (EdgeConv), generating 64-dimensional embeddings. The four embeddings are combined via 1D convolution and passed through fully connected layers for final classification.}
    \label{fig:model_architecture}
\end{figure*}

The KIGNet architecture processes the four graph representations in parallel through identical but independently parameterized networks, then combines their outputs for final classification. Figure~\ref{fig:model_architecture} illustrates the complete pipeline.

\subsubsection{Feature Extraction with Hybrid Convolutional Layers}

Each of the four graphs passes through its own feature extractor consisting of five graph convolutional layers. We alternate between two complementary types of graph convolutions to capture both local geometric structure and broader graph-level patterns.

\paragraph{Chebyshev Graph Convolutions.}
A general graph convolutional operator is defined as $x' = p_w(L)\,x$, where $x$ is the node feature matrix of the graph and $p_w(L)$ is a matrix polynomial of the form:
\begin{equation}
\label{eq:poly_operator}
p_w(L) = \sum_{i=0}^{d} w_i \, L^i
\end{equation}
where $w_i$ are learnable scalar parameters. The fully expanded operator for node $v$ is:
\begin{equation}
\label{eq:expanded_operator}
x'_v = \sum_{i=0}^{d} w_i \sum_{u \in V} [L^i]_{u,v} \, x_u
\end{equation}
where $V$ is the full node set and $[L^i]_{u,v}$ denotes the $(u,v)$ entry of the matrix power $L^i$. Because $L$ is sparse, $[L^i]_{u,v} = 0$ for nodes $u$ and $v$ separated by more than $i$ hops, so the effective receptive field of an order-$i$ filter is the $i$-hop neighborhood of $v$; the sum over all $u \in V$ is therefore equivalent to summing over the $i$-hop neighborhood without artificially restricting the formula.

The Chebyshev convolutional operator replaces the raw Laplacian 
powers with Chebyshev polynomial filters~\cite{Defferrard2016_Chebyshev, HAMMOND2011_GraphSpectral} applied to the normalized 
Laplacian:
\begin{equation}
\label{eq:cheb_operator}
p_w(L) = \sum_{i=0}^{d} w_i \, T_i(L_{\text{norm}})
\end{equation}
yielding the full Chebyshev convolutional operator for node $v$:
\begin{equation}
\label{eq:cheb_node}
x'_v = \sum_{i=0}^{d} w_i \sum_{u \in V} 
    [T_i(L_{\text{norm}})]_{u,v} \, x_u
\end{equation}
where $[T_i(L_{\text{norm}})]_{u,v}$ is the $(u,v)$ entry of the 
$i$-th Chebyshev matrix polynomial evaluated at $L_{\text{norm}}$, 
$T_i$ is the $i$-th Chebyshev polynomial with initial conditions 
$T_0(x) = 1$, $T_1(x) = x$ and recurrence:
\begin{equation}
\label{eq:cheb_recurrence}
T_i(x) = 2x \; T_{i-1}(x) - T_{i-2}(x)
\end{equation}
and $L_{\text{norm}}$ is the normalized graph Laplacian:
\begin{equation}
\label{eq:laplacian_norm}
L_{\text{norm}} = \frac{2L}{\lambda_{\max}} - I
\end{equation}
where $\lambda_{\max}$ is the largest eigenvalue of $L$ and $I$ is 
the identity matrix. Chebyshev polynomials provide computational 
efficiency by approximating spectral graph filters without fully 
diagonalizing the Laplacian~\cite{Shuman2011_Chebyshev_polynomial, Shuman2013emerging}, which becomes infeasible at large 
scales.

\paragraph{Advantages of Chebyshev Convolutions.}
ChebConv outperforms classical graph convolutions for two main reasons. Classical graph convolutions multiply a sum of neighbor node features by a learnable weight matrix, which can lead to oversmoothing that loses fine-grained detail, and repeated convolutions may cause vanishing gradients. ChebConv addresses both issues by applying Chebyshev polynomial filters that focus on a small neighborhood around each node, preserving local graph structure and mitigating oversmoothing. This is particularly beneficial for jets that exhibit patterns at various spatial resolutions. ChebConv also adapts to graphs of different sizes by operating on graph structure rather than relying on a fixed-size filter, which is advantageous when processing jets with varying numbers of constituent particles.

\paragraph{Edge Convolutions.}
Edge convolutions (EdgeConv)~\cite{Wang2019_DGCNN} complement Chebyshev convolutions by focusing on pairwise relationships between connected particles. EdgeConv uses two independently learnable sets of parameters to encode both global shape and local relational information:
\begin{equation}
\label{eq:edgeconv}
x'_i = \max_{j \in \mathcal{N}(i)} \!\left[\Theta \cdot (x_i - x_j)\right] 
       + \Phi \cdot x_i
\end{equation}
where $\Theta$ and $\Phi$ are independently learnable weight matrices. The max-aggregation applies only to the relative term $\Theta \cdot (x_i - x_j)$, which encodes the asymmetric relationship between particle $i$ and each neighbor $j$; the self-term $\Phi \cdot x_i$ is independent of the neighborhood and is added after aggregation, encoding the global shape contribution of particle $i$ itself. EdgeConv excels at capturing the broader relational context within a jet, analyzing how localized interactions aggregate into distinct jet features observable at larger scales.

\subsubsection{Synthesis of Local and Global Features}

The decision to interleave EdgeConv layers between ChebConv layers is grounded in the rationale that this configuration facilitates the concurrent extraction of local features (via ChebConv) and relational information (via EdgeConv) at each processing stage. A purely sequential arrangement, all ChebConv followed by all EdgeConv, would defer the integration of global context until the later layers. Given the intricate, multi-resolution structures of jets, an interleaved design is preferred for comprehensive feature extraction.

The interleaved structure also accommodates jets with varying numbers of constituent particles. ChebConv's adaptability to different graph sizes is enhanced by the intervening EdgeConv layers, which allow the model to dynamically adjust its receptive field rather than progressively extending to larger neighborhoods through consecutive ChebConv layers alone.

The alternating pattern (ChebConv $\to$ EdgeConv $\to$ ChebConv $\to$ EdgeConv $\to$ ChebConv) is followed by batch normalization (BN) and ReLU activation after each layer. The initial layer transforms the 16-dimensional input features to 64-dimensional hidden representations, which are maintained through subsequent layers.

\subsubsection{Graph-Level Pooling and Classification}

\paragraph{Graph-Level Pooling.} After the final convolutional layer, all particle-level features are aggregated into a single 64-dimensional vector representing the entire jet via mean pooling. This produces four embedding vectors per jet, one from each graph branch.

\paragraph{Multi-Graph Combination.} The four 64-dimensional vectors are stacked into a $4 \times 64$ matrix. A 1D convolution with kernel size 1 is applied across the four graph channels, learning how to weight and combine information from the different kinematic representations.

\paragraph{Classification.} The combined representation is flattened to a 256-dimensional vector and passed through two fully connected (FC) layers with dropout rate 0.1 for regularization. The first layer reduces dimensionality to 64, and the second produces 10 outputs. A softmax function converts these outputs to class probabilities (Algorithm~\ref{alg:forward_pass}).

\begin{algorithm}[htbp]
\caption{KIGNet Forward Pass}
\label{alg:forward_pass}
\begin{algorithmic}[1]
\Require Weighted graphs $G_{\Delta}, G_{k_T}, G_z, G_{m^2}$ (16-dim nodes)
\Ensure Class probabilities over 10 classes
\For{each graph $G \in \{G_{\Delta}, G_{k_T}, G_z, G_{m^2}\}$}
    \State Apply 5 alternating ChebConv-EdgeConv layers to $G$ (ChebConv first; $16{\to}64$; BN, ReLU after each)
    \State $\mathbf{e}_G \gets$ mean-pool the result over particles \Comment{$64$-dim}

\EndFor
\State Stack $\{\mathbf{e}_G\}$ into a $4{\times}64$ matrix; combine branches via $1{\times}1$ conv ($+$ BN, ReLU)
\State Flatten to 256; FC $256{\to}64{\to}10$ (BN, ReLU, dropout $0.1$ after the first)
\State Apply softmax over the 10 outputs
\State \Return class probabilities
\end{algorithmic}
\end{algorithm}

\subsection{Training Configuration}
\label{subsec:training_config}

\paragraph{Hyperparameters.} Table~\ref{tab:hyperparameters} details the hyperparameters used for KIGNet training and architecture.

\begin{table}[htbp]
\centering
\caption{Hyperparameters for KIGNet training and architecture.}
\label{tab:hyperparameters}
\begin{tabular*}{\columnwidth}{@{\extracolsep{\fill}}ll|lr}
\toprule \toprule
\multicolumn{2}{c|}{Optimization} & \multicolumn{2}{c}{Architecture} \\
\midrule
Optimizer & AdamW & Hidden Dim. & 64 \\
Learning Rate & $10^{-3}$ & Graph Branches & 4 \\
Batch Size & 256 &  $k$-NN & 3 \\
Max Epochs & 500 &  Convolution Layers & 5\\
Convergence Threshold & $10^{-4}$ &  Output Classes & 10\\
Early Stop & 10 epochs & & \\
\midrule
\multicolumn{4}{c}{Regularization} \\
\multicolumn{2}{c|}{Dropout Rate} & \multicolumn{2}{c}{0.1} \\
\bottomrule \bottomrule
\end{tabular*}
\end{table}

\paragraph{Training Procedure.} Jets are loaded in batches of 256 with random shuffling at each epoch. For each jet, the four graphs ($G_{\Delta}$, $G_{k_T}$, $G_{z}$, $G_{m^2}$) are constructed on-the-fly using $k$-NN connectivity based on angular distance in $(\eta, \phi)$ space. Each graph is independently processed through its respective feature extractor, yielding four 64-dimensional embeddings subsequently combined via 1D convolution. Cross-entropy loss is minimized, and network parameters are updated via the AdamW optimizer with learning rate $10^{-3}$. Training employs early stopping: termination occurs when validation loss shows no improvement exceeding $10^{-4}$ over 10 consecutive epochs.

\subsection{Kinematic Variable Importance via Grad-CAM}
\label{sec:gradcam}

Before describing the explainability methodology, we note an important interpretive constraint: the JetClass dataset comprises simulated jets generated using Pythia 8.230~\cite{Sjostrand2015}, which implements QCD-based parton showers, matrix element calculations, and hadronization through the Lund string model~\cite{Andersson1983}. The kinematic variables we analyze, $\Delta$, $k_T$, $z$, and $m^2$, directly correspond to coordinates in the Lund jet plane~\cite{Dreyer2018_LundJetPlane}, a representation explicitly designed to expose QCD splitting function structure. Our explainability analysis therefore measures whether KIGNet learns to exploit the QCD factorization structure present in the training data. Agreement between learned feature importance and theoretical QCD predictions confirms that the model captures physically motivated patterns, but does not provide independent experimental validation, as both the model and the theory operate on the same simulated data substrate.

We adapt the concept of Gradient-weighted Class Activation Mapping (Grad-CAM)~\cite{Selvaraju2017_GradCAM}, a technique originally developed for interpreting convolutional neural networks, to our multi-graph architecture. Our model processes four distinct graph types through dedicated feature extractors, each producing 64-dimensional embeddings. For each graph type, we compute the gradient of the predicted class score with respect to its corresponding embedding to measure variable importance. The importance of each graph type is calculated as the product of the gradient magnitude and the embedding magnitude, globally averaged across the 64 embedding dimensions, yielding a scalar importance score per graph type. We compute importance scores independently for each test sample and each jet type, then average across all test samples for global importance and within each jet type for class-specific importance. The four raw importance scores are normalized to percentages summing to 100\% to facilitate comparison. The detailed procedure is outlined in Algorithm~\ref{alg:gradcam}.

\begin{algorithm}[htbp]
\caption{GradCAM: Compute Graph Type Importance}
\label{alg:gradcam}
\begin{algorithmic}[1]
\Require Branch graphs $G_{\Delta}, G_{k_T}, G_z, G_{m^2}$ of one jet; target class $c$ (true label, else predicted)
\Ensure Per-branch importances (\%, summing to 100)
\For{each branch $g \in \{\Delta, k_T, z, m^2\}$}
    \State $\mathbf{e}_g \gets \mathrm{model}_g(G_g)$ \Comment{$64$-dim}
\EndFor
\State $s \gets$ class-$c$ logit of the classifier on stacked $\{\mathbf{e}_g\}$
\State Backpropagate $s$ to get $\nabla_{\mathbf{e}_g}$ for each $g$
\For{each branch $g \in \{\Delta, k_T, z, m^2\}$}
    \State $\mathrm{imp}_g \gets \mathrm{Mean}\!\left(\lvert\nabla_{\mathbf{e}_g}\rvert \times \lvert\mathbf{e}_g\rvert\right)$ \Comment{gradient $\times$ activation}
\EndFor
\State Normalize $\{\mathrm{imp}_g\}$ to percentages (sum $=100\%$)
\State \Return $\{\mathrm{imp}_g\}$
\end{algorithmic}
\end{algorithm}

\section{Results} 
\label{sec:results}

\begin{table*}[htbp]
\centering
\caption{Per-class performance of KIGNet and the PCN baseline on the 20M-jet test set of the \texttt{JetClass} dataset. For the macro-average and each of the signal classes, we report classification accuracy, AUC, and AUPR, together with the relative improvement (Imp.\%) of KIGNet over the PCN baseline.}
\label{tab:jetclass_performance}
\begin{tabular*}{\textwidth}{@{\extracolsep{\fill}}lccccccccc}
\toprule \toprule
& \multicolumn{3}{c}{Accuracy} & \multicolumn{3}{c}{AUC} & \multicolumn{3}{c}{AUPR} \\
\cmidrule(lr){2-4} \cmidrule(lr){5-7} \cmidrule(lr){8-10}
Class & PCN & KIGNet (ours) & Imp.\% & PCN & KIGNet (ours) & Imp.\% & PCN & KIGNet (ours) & Imp.\% \\
\midrule
Macro-Avg & 0.9280 & \textbf{0.9507} & +2.45 & 0.9343 & \textbf{0.9661} & +3.40 & 0.6844 & \textbf{0.8152} & +19.11 \\
\midrule
$H \rightarrow b\bar{b}$ & 0.9045 & \textbf{0.9564} & +5.74 & 0.8781 & \textbf{0.9761} & +11.16 & 0.4738 & \textbf{0.8601} & +81.53 \\
$H \rightarrow c\bar{c}$ & 0.8992 & \textbf{0.9300} & +3.43 & 0.8633 & \textbf{0.9287} & +7.58 & 0.4577 & \textbf{0.6936} & +51.54 \\
$H \rightarrow gg$ & 0.8986 & \textbf{0.9313} & +3.64 & 0.9190 & \textbf{0.9501} & +3.38 & 0.5741 & \textbf{0.7242} & +26.15 \\
$H \rightarrow 4q$ & 0.9089 & \textbf{0.9325} & +2.60 & 0.9362 & \textbf{0.9570} & +2.22 & 0.6431 & \textbf{0.7266} & +12.98 \\
$H \rightarrow \ell\nu qq'$ & 0.9635 & \textbf{0.9815} & +1.87 & 0.9825 & \textbf{0.9941} & +1.18 & 0.8999 & \textbf{0.9658} & +7.32 \\
$t \rightarrow bqq'$ & 0.9483 & \textbf{0.9677} & +2.05 & 0.9738 & \textbf{0.9872} & +1.38 & 0.8043 & \textbf{0.9085} & +12.96 \\
$t \rightarrow b\ell\nu$ & 0.9773 & \textbf{0.9878} & +1.07 & 0.9929 & \textbf{0.9975} & +0.46 & 0.9532 & \textbf{0.9837} & +3.20 \\
$W \rightarrow qq'$ & 0.9182 & \textbf{0.9248} & +0.72 & 0.9313 & \textbf{0.9507} & +2.08 & 0.6024 & \textbf{0.6694} & +11.12 \\
$Z \rightarrow q\bar{q}$ & 0.9331 & \textbf{0.9444} & +1.21 & 0.9316 & \textbf{0.9531} & +2.31 & 0.7513 & \textbf{0.8047} & +7.11 \\
\bottomrule \bottomrule
\end{tabular*}
\end{table*}

This section evaluates KIGNet on the \texttt{JetClass} dataset. We report classification accuracy in Section~\ref{sec:classification_accuracy}, AUC metrics in Section~\ref{sec:auc}, and AUPR in Section~\ref{sec:aupr}. We compare KIGNet with existing state-of-the-art models in Section~\ref{subsec:sota_comparison} and provide a Grad-CAM explainability analysis in Section~\ref{subsec:explainability_result}.

\subsection{Classification Accuracy}
\label{sec:classification_accuracy}

Table~\ref{tab:jetclass_performance} presents classification accuracies for all signal classes. KIGNet achieves a macro-averaged accuracy of 0.9507, a 2.45\% improvement over PCN\footnote{PCN denotes our implementation of the baseline Particle Chebyshev Network trained on the same 1M jets (100K per class) and tested on 20M jets from the \texttt{JetClass} dataset, using the same training and test split as KIGNet (Table~\ref{tab:data_split}).} (0.9280), outperforming the baseline across all nine signal classes. The gains are largest for the Higgs decay channels, led by $H \rightarrow b\bar{b}$ at 5.74\% (0.9564 vs.\ 0.9045), followed by $H \rightarrow gg$ at 3.64\% and $H \rightarrow c\bar{c}$ at 3.43\%. Improvements are smaller but still consistent for the electroweak boson and top-quark channels, ranging from 0.72\% for $W \rightarrow qq'$ to 2.60\% for $H \rightarrow 4q$; the full per-class breakdown is given in Table~\ref{tab:jetclass_performance}.

\subsection{Area Under the Curve (AUC)}
\label{sec:auc}

Table~\ref{tab:jetclass_performance} compares AUC performance. KIGNet achieves a macro-averaged AUC of 0.9661, a 3.40\% improvement over PCN (0.9343). The largest gains again occur in the heavy-flavor channels, where $H \rightarrow b\bar{b}$ improves by 11.16\% (0.8781 to 0.9761) and $H \rightarrow c\bar{c}$ by 7.58\% (0.8633 to 0.9287), reflecting the difficulty of separating these classes under the PCN baseline. Top-quark decays show the smallest relative gains ($t \rightarrow bqq'$: +1.38\%; $t \rightarrow b\ell\nu$: +0.46\%), as the PCN baseline already approaches saturation in this regime (AUC $>0.97$); electroweak boson decays improve by roughly 2\% for both $W \rightarrow qq'$ and $Z \rightarrow q\bar{q}$.

\subsection{Area Under the Precision-Recall Curve (AUPR)}
\label{sec:aupr}

Table~\ref{tab:jetclass_performance} details AUPR performance, a metric particularly informative for imbalanced datasets. KIGNet achieves a macro-averaged AUPR of 0.8152, a 19.11\% improvement over PCN (0.6844). The improvement is concentrated almost entirely in the heavy-flavor channels: $H \rightarrow b\bar{b}$ improves by 81.53\% (0.4738 to 0.8601) and $H \rightarrow c\bar{c}$ by 51.54\% (0.4577 to 0.6936), with $H \rightarrow gg$ and $H \rightarrow 4q$ improving by 26.15\% and 12.98\%, respectively. This pattern, large AUPR gains specifically for $b$- and $c$-initiated jets, is consistent with the heightened sensitivity of the $m^2$ graph branch to heavy-quark dead-cone effects discussed in Section~\ref{sec:qcd_consistency}. Gains for the remaining channels are more modest, ranging from 3.20\% ($t \rightarrow b\ell\nu$) to 12.96\% ($t \rightarrow bqq'$); see Table~\ref{tab:jetclass_performance} for the complete set of values.

\subsection{Comparison to State-of-the-Art Models}
\label{subsec:sota_comparison}

We compare KIGNet against five baseline models trained on \texttt{JetClass}: PCN, PFN~\cite{Komiske2019_EFN}, P-CNN~\cite{cms2017boosted}, ParticleNet~\cite{Qu2020_ParticleNet}, and ParT~\cite{Qu2022_ParticleTransformer}. Table~\ref{tab:sota_comparison} summarizes performance metrics.

\begin{table}[htbp]
\centering
\caption{Comparison of model performances on the \texttt{JetClass} dataset. Percentages in parentheses indicate the relative improvement of KIGNet over the corresponding method.}
\label{tab:sota_comparison}
\begin{tabular*}{\columnwidth}{@{\extracolsep{\fill}}l@{\hspace{10pt}}rr@{\hspace{20pt}}rr}
\toprule \toprule
Model & \multicolumn{2}{c}{Macro-Accuracy}  & \multicolumn{2}{c}{Macro-AUC}  \\
\midrule
PFN~\cite{Komiske2019_EFN} & 0.772 & (+23.2\%) & 0.9714 & ($-$0.5\%) \\
P-CNN~\cite{cms2017boosted} & 0.809 & (+17.6\%) & 0.9789 & ($-$1.3\%) \\
ParticleNet~\cite{Qu2020_ParticleNet} & 0.844 & (+12.7\%) & 0.9849 & ($-$1.9\%) \\
ParT~\cite{Qu2022_ParticleTransformer} & 0.861 & (+10.5\%) & \textbf{0.9877} & ($-$2.2\%) \\
\midrule
PCN (baseline) & 0.928 & (+2.5\%) & 0.9343 & (+3.4\%) \\
KIGNet (ours) & \multicolumn{2}{c}{\textbf{0.951}} & \multicolumn{2}{c}{0.9661} \\
\bottomrule \bottomrule
\end{tabular*}
\end{table}

KIGNet sets a new benchmark for classification accuracy, achieving 0.9507 and outperforming the state-of-the-art ParT model (0.861) by 10.5\%. In terms of AUC, KIGNet (0.9661) performs comparably to ParT (0.9877), trailing by 2.2 percentage points while maintaining the accuracy advantage.

The model's effectiveness stems from its multi-graph architecture, which processes kinematic variables ($\Delta$, $k_T$, $z$, $m^2$) in parallel branches, capturing complementary physical information of angular structure, momentum scales, energy sharing, and mass relationships simultaneously. A 1D convolution module dynamically weights these representations using explicit Lund plane features~\cite{Dreyer2018_LundJetPlane}. While this multi-branch design increases computational overhead compared to single-graph models, it produces substantial gains in accuracy and precision-recall metrics.

\subsection{Explainability Analysis}
\label{subsec:explainability_result}

We apply the Grad-CAM concept to understand which kinematic variables the network prioritizes when classifying jets. The method quantifies how sensitive the network's output is to each graph representation: if modifying a particular graph branch substantially affects the prediction, that feature is deemed important.

Table~\ref{tab:classwise_importance} presents the class-wise feature importance. Angular separation ($\Delta$) demonstrates the highest contribution for most classes, ranging from 38.27\% to 45.05\%, with particularly high importance for leptonic channels ($H \to \ell\nu qq'$ at 45.05\% and $t \to b\ell\nu$ at 43.98\%). Relative transverse momentum ($k_T$) shows complementary importance, contributing most strongly to gluon jets ($H \to gg$ at 40.24\%) and hadronic top decays ($t \to bqq'$ at 38.74\%). Together, $\Delta$ and $k_T$ account for approximately 76\% of the classification signal across all classes. Momentum fraction ($z$) maintains moderate importance (12.22\%-15.54\%), while invariant mass squared ($m^2$) shows the lowest overall contribution (7.68\%-13.64\%). Notably, $m^2$ exhibits higher relevance for heavy-flavor jets such as $H \to b\bar{b}$ (13.64\%), consistent with the theoretical expectation that invariant mass provides discrimination for heavy-quark identification.

\begin{table}[htbp]
\centering
\caption{Grad-CAM feature importance for jet classification. Values represent the relative contribution (\%) of each kinematic variable (angular separation $\Delta$, relative transverse momentum $k_T$, momentum fraction $z$, and invariant mass squared $m^2$) to the KIGNet classification decision for each jet class. Bold values indicate the most important feature per class. Angular separation and relative transverse momentum consistently dominate, together contributing approximately 76\% of the feature importance.}
\label{tab:classwise_importance}
\begin{tabular*}{\columnwidth}{@{\extracolsep{\fill}}lcccc}
\toprule \toprule
Class & $\Delta$ & $k_T$ & $z$ & $m^2$ \\
\midrule
$H \rightarrow b\bar{b}$  & \textbf{38.27} & 35.59          & 12.49 & 13.64 \\
$H \rightarrow c\bar{c}$  & \textbf{39.98} & 36.40          & 13.93 & 09.68 \\
$H \rightarrow gg$        & 38.37          & \textbf{40.24} & 13.30 & 08.09 \\
$H \rightarrow 4q$        & \textbf{39.13} & 38.93          & 14.26 & 07.68 \\
$H \rightarrow \ell\nu qq'$     & \textbf{45.05} & 31.72          & 14.55 & 08.68 \\
$t \rightarrow bqq'$      & 38.32          & \textbf{38.74} & 12.22 & 10.72 \\
$t \rightarrow b\ell\nu$       & \textbf{43.98} & 30.60          & 14.77 & 10.66 \\
$W \rightarrow qq'$       & \textbf{39.19} & 37.44          & 15.36 & 08.01 \\
$Z \rightarrow q\bar{q}$  & \textbf{42.61} & 33.61          & 15.54 & 08.24 \\
\bottomrule \bottomrule
\end{tabular*}
\end{table}

\begin{figure*}[htbp]
    \centering
    \includegraphics[width=\textwidth]{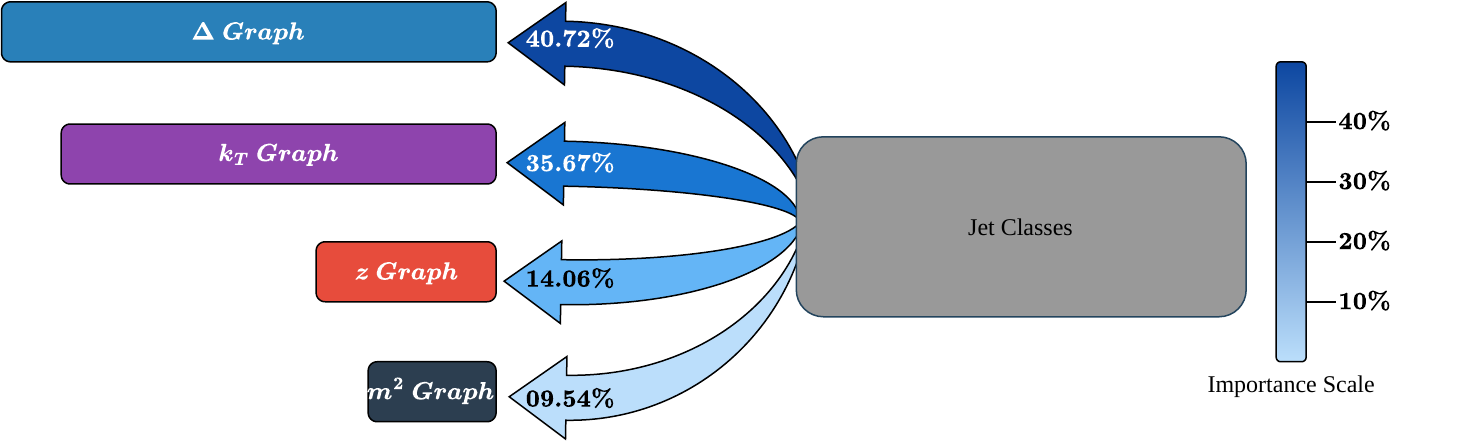}
    \caption{Grad-CAM feature importance for KIGNet classification. (a) Four graph representations are processed through parallel GNN branches. (b) Angular separation ($\Delta$, 40.72\%) and transverse momentum ($k_{T}$, 35.67\%) dominate classification, while momentum fraction ($z$, 14.06\%) and invariant mass squared ($m^{2}$, 9.54\%) provide complementary information (Algorithm~\ref{alg:gradcam}). This hierarchy is consistent with QCD factorization in the Lund plane (Eq.~\eqref{eq:emission_full}), and class-specific variations (Table~\ref{tab:classwise_importance}) reflect distinct QCD mechanisms: enhanced $k_T$ for gluon radiation, elevated $\Delta$ for leptonic decays, and increased $m^2$ for heavy-quark dead-cone effects.}
    \label{fig:gradcam}
\end{figure*}

Figure~\ref{fig:gradcam} presents Grad-CAM results averaged over the test dataset. The analysis reveals a clear hierarchy: $\Delta$ accounts for 40.72\% of the decision making weight, followed by $k_T$ at 35.67\%, with $z$ and $m^2$ contributing 14.06\% and 9.54\%, respectively (Table~\ref{tab:feature_importance}).

\begin{table}[htbp]
\centering
\caption{Global feature importance from Grad-CAM analysis. Angular separation ($\Delta$) and transverse momentum ($k_T$) dominate classification decisions, while momentum fraction ($z$) and invariant mass squared ($m^2$) provide complementary information.}
\label{tab:feature_importance}
\begin{tabular*}{\columnwidth}{@{\extracolsep{\fill}}p{1.5cm}cp{4.5cm}}
\toprule \toprule
Feature Graph & Importance (\%) & Description \\
\midrule
$\Delta$ Graph & 40.72\% & Angular separation between particles; dominant feature in classification. \\
$k_T$ Graph & 35.67\% & Relative transverse momentum scale associated with jet splitting. \\
$z$ Graph & 14.06\% & Momentum fraction carried by particles in the jet. \\
$m^2$ Graph & 9.54\% & Invariant mass squared of particle pairs, providing complementary structural information. \\
\bottomrule \bottomrule
\end{tabular*}
\end{table}

The observed feature importance hierarchy aligns with QCD expectations in a manner that requires careful interpretation. The dominance of $\Delta$ and $k_T$ reflects the soft-collinear factorization structure of perturbative QCD, where emission probabilities factorize as in Eq.~\eqref{eq:emission_full}~\cite{Dreyer2018_LundJetPlane}. However, this agreement is expected rather than surprising: JetClass jets are generated using Pythia 8, which implements these same QCD splitting functions. Our analysis therefore demonstrates that KIGNet successfully learns to exploit the QCD structure encoded in the training data, a validation of the architecture's ability to discover physically motivated patterns, but does not constitute independent experimental confirmation of QCD. The consistency between learned importance and QCD predictions nevertheless suggests the model relies on robust physical correlations rather than simulation-specific artifacts, providing confidence for deployment on experimental data. Ultimate validation requires evaluation on real collider measurements with full detector effects, pileup, and systematic uncertainties, which we begin to explore in Section~\ref{subsec:aspen_results}.

\subsection{Generalization to Real Data: Aspen Open Jets}
\label{subsec:aspen_results}

We evaluate KIGNet on the Aspen Open Jets (AOJ) dataset~\cite{Amram2024AOJ}, which comprises approximately 178 million high-$p_T$ jets extracted from the CMS 2016 Open Data. Unlike JetClass, AOJ includes detector effects, pileup contamination, and reconstruction uncertainties, providing a realistic test of whether the model captures physically meaningful features rather than simulation-specific correlations.

\begin{figure*}[htbp]
    \centering
    \includegraphics[width=\textwidth]{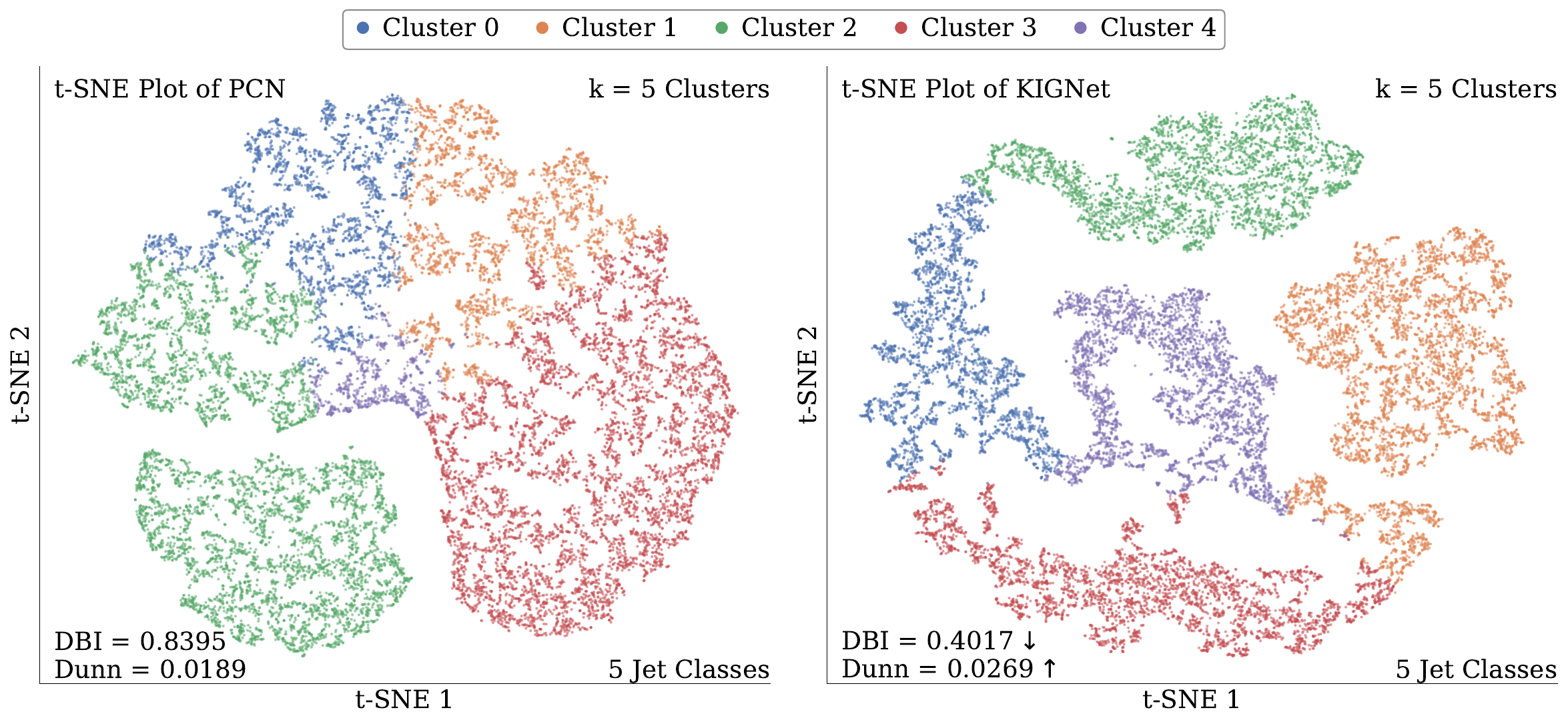}
    \caption{Two-dimensional t-SNE projections of jet embeddings on the Aspen Open Jets dataset (Algorithm~\ref{alg:tsne_viz}, perplexity $=30$). Cluster quality is summarized by the Davies-Bouldin Index (DBI $\downarrow$, lower is better) and the Dunn Index ($\uparrow$, higher is better). PCN embeddings (left; DBI $=0.8395$, Dunn $=0.0189$) are diffuse and overlapping, indicating limited separability, whereas KIGNet (right; DBI $=0.4017$ $\downarrow$, Dunn $=0.0269$ $\uparrow$) produces substantially more compact and well-separated clusters, confirming that explicit encoding of Lund plane kinematic variables yields more structured latent representations on real CMS collision data subject to detector effects, pileup, and reconstruction uncertainties.}
    \label{fig:tsne}
\end{figure*}

\paragraph{Clustering Quality Evaluation.}
Since clean multi-class labels comparable to JetClass are not publicly available for AOJ, we assess representation quality using unsupervised clustering metrics applied to the learned jet embeddings. The procedure is described in Algorithm~\ref{alg:clustering_eval}: embeddings are extracted from the frozen model, clustered via $k$-means, and evaluated using the Davies-Bouldin Index (DBI) and the Dunn Index.

\paragraph{Unsupervised Training via DeepCluster.}
Because the AOJ dataset does not provide ground-truth class labels, we train the KIGNet encoder in a fully unsupervised manner using the DeepCluster algorithm~\cite{Caron_2018_ECCV}, originally proposed for self-supervised representation learning on natural images. DeepCluster alternates between two steps: (i) extracting feature embeddings with the current encoder and assigning pseudo-labels to each sample via $k$-means clustering in the embedding space, and (ii) updating the encoder parameters by treating these pseudo-labels as targets in a standard cross-entropy classification objective with a lightweight linear head, optimized with stochastic gradient descent (SGD). The head is re-initialized at every iteration to prevent the network from memorizing spurious label assignments, and empty or near-empty clusters are reassigned by splitting the largest cluster with a small perturbation to preserve diversity. This cycle of \emph{cluster, label, and train} drives the encoder to produce embeddings that are increasingly structured and cluster-friendly without requiring any manual annotation. We adapt the original image-domain procedure of Caron et al.~\cite{Caron_2018_ECCV} to the multi-graph jet setting by replacing the convolutional image backbone with our four-branch KIGNet encoder, L2-normalizing embeddings before clustering, and using class-balanced cross-entropy weights to mitigate cluster-size imbalance. Convergence is monitored via the Davies-Bouldin Index on the training embeddings together with the normalized mutual information (NMI) between consecutive pseudo-label assignments; the full protocol is given in Algorithm~\ref{alg:deepcluster_train}, and the post-hoc clustering quality evaluation on the test set is described in Algorithm~\ref{alg:clustering_eval}. For a fair architectural comparison, we apply the identical DeepCluster protocol (Algorithm~\ref{alg:deepcluster_train}) to both the PCN baseline and KIGNet encoders, holding the optimizer, learning-rate schedule, and all other training hyperparameters fixed; the only difference between the two runs is the encoder architecture.

\begin{algorithm}[htbp]
\caption{DeepCluster Training of the Encoder on AOJ}
\label{alg:deepcluster_train}
\begin{algorithmic}[1]
\Require AOJ jet graphs $\{G_i\}$; $K$ clusters; max iterations $T$; random-init encoder $f_\theta$
\Ensure Best encoder $f_{\theta^*}$ (lowest training DBI); final pseudo-labels $\{c_i\}$
\State $c^{\text{prev}} \gets \varnothing$; $\mathrm{DBI}^{\text{best}} \gets \infty$; $f_{\theta^*} \gets f_\theta$
\For{$t = 1$ to $T$}
    \State $\mathbf{e}_i \gets$ L2-normalize $f_\theta(G_i)$, for all jets
    \State $\{c_i\} \gets k\text{-means}(\{\mathbf{e}_i\}, K)$ \Comment{empty clusters reseeded; see Sec.~\ref{subsec:aspen_results}}
    \State $\mathrm{DBI}_t \gets \mathrm{DBI}(\{\mathbf{e}_i\}, \{c_i\})$ \Comment{Eq.~\eqref{eq:dbi}}
    \If{$c^{\text{prev}} \neq \varnothing$ and $\mathrm{NMI}(c^{\text{prev}}, \{c_i\}) > 0.99$ and $t > 10$}
        \State \textbf{break} \Comment{pseudo-labels stabilized}
    \EndIf
    \State $c^{\text{prev}} \gets \{c_i\}$
    \State Train $f_\theta$ + reinitialized linear head for one epoch to predict $\{c_i\}$ \Comment{class-balanced CE; SGD, see Sec.~\ref{subsec:aspen_results}}
    \If{$\mathrm{DBI}_t < \mathrm{DBI}^{\text{best}}$}
        \State $\mathrm{DBI}^{\text{best}} \gets \mathrm{DBI}_t$; \ $f_{\theta^*} \gets f_\theta$
    \EndIf
\EndFor
\State \Return $f_{\theta^*}$, $\{c_i\}$
\end{algorithmic}
\end{algorithm}

\begin{algorithm}[htbp]
\caption{AOJ Test-Set Clustering Evaluation}
\label{alg:clustering_eval}
\begin{algorithmic}[1]
\Require Trained encoder $f_{\theta^*}$; AOJ training/test graphs; $K$ clusters
\Ensure Davies-Bouldin Index (lower is better); Dunn Index (higher is better)
\State Encode and L2-normalize all training/test jets with frozen $f_{\theta^*}$
\State $\{\boldsymbol{\mu}_k\} \gets k\text{-means}(K)$ on training embeddings
\State Assign each test jet to its nearest centroid, giving $\{c_j\}$
\State Compute DBI from $\{\mathbf{e}_j\}, \{c_j\}, \{\boldsymbol{\mu}_k\}$ \Comment{Eq.~\eqref{eq:dbi}}
\State Compute Dunn Index from $\{\mathbf{e}_j\}, \{c_j\}$ \Comment{Eq.~\eqref{eq:dunn}}
\State \Return DBI, Dunn
\end{algorithmic}
\end{algorithm}

\begin{algorithm}[htbp]
\caption{t-SNE Visualization of Test Embeddings}
\label{alg:tsne_viz}
\begin{algorithmic}[1]
\Require Test embeddings $\{\mathbf{e}_j\}$; cluster labels $\{c_j\}$ (from Algorithm~\ref{alg:clustering_eval}); perplexity $\mathcal{P}$
\Ensure 2D scatter plot colored by $\{c_j\}$
\State Subsample $\le 20{,}000$ test embeddings uniformly at random
\State Project to 2D via t-SNE ($\mathcal{P} = 30$, seed $42$) \Comment{Eqs. \eqref{eq:tsne_pji} - \eqref{eq:tsne_kl}}
\State Plot projected points, colored by $\{c_j\}$
\Statex \quad t-SNE preserves local structure only; 2D inter-cluster distances are not quantitative.
\State \Return scatter plot
\end{algorithmic}
\end{algorithm}

Table~\ref{tab:aspen_clustering} reports the Davies-Bouldin Index (DBI) and Dunn Index for both PCN and KIGNet.

\begin{table}[htbp]
\centering
\caption{Clustering quality of PCN and KIGNet embeddings on the Aspen Open Jets real CMS collision dataset. Lower DBI indicates more compact clusters; higher Dunn Index indicates better inter-cluster separation. KIGNet reduces DBI by 52.15\% and increases the Dunn Index by 42.33\%.}
\label{tab:aspen_clustering}
\begin{tabular*}{\columnwidth}{@{\extracolsep{\fill}}lcc}
\toprule \toprule
Model & Davies-Bouldin Index $\downarrow$ & Dunn Index $\uparrow$ \\
\midrule
PCN   & $0.8395$ & $0.0189$ \\
KIGNet (ours) & $\boldsymbol{0.4017}$ $\boldsymbol{\downarrow}$ & $\boldsymbol{0.0269}$ $\boldsymbol{\uparrow}$ \\
\midrule
Improvement (\%) & $+52.15\%$ $\downarrow$ & $+42.33\%$ $\uparrow$\\
\bottomrule \bottomrule
\end{tabular*}
\end{table}

KIGNet reduces the Davies-Bouldin Index by approximately 52.15\% relative to PCN, indicating significantly improved cluster compactness and inter-cluster separation. Simultaneously, the Dunn Index increases by 42.33\%, reflecting a substantial enhancement in the minimum inter-cluster distance relative to intra-cluster spread.

\paragraph{Embedding Space Visualization.}
To complement the quantitative clustering metrics, we project the 64-dimensional jet embeddings into two dimensions using $t$-distributed Stochastic Neighbor Embedding (t-SNE)~\cite{tsne_maaten}, a nonlinear dimensionality reduction technique that preserves local neighborhood structure. The full procedure is described in Algorithm~\ref{alg:tsne_viz}.

The key hyperparameters are perplexity $\mathcal{P}$, which controls the effective number of neighbors considered for each point and is set to 30 in our experiments, and the Student-$t$ kernel in the low-dimensional space, which places heavier mass on distant points than a Gaussian and thereby prevents cluster collapse. Embeddings are colored by the $k$-means cluster assignments obtained from Algorithm~\ref{alg:clustering_eval}, so the visual structure directly corresponds to the quantitative DBI and Dunn Index values. It is important to note that t-SNE preserves local neighborhoods but distorts global distances; inter-cluster distances in the 2D projection should therefore not be interpreted quantitatively, and the clustering metrics in Table~\ref{tab:aspen_clustering} remain the primary quantitative evidence.

Figure~\ref{fig:tsne} presents the resulting two-dimensional projections. PCN baseline embeddings exhibit diffuse and overlapping clusters, indicating limited separability between underlying jet populations. In contrast, KIGNet produces more compact and well-separated clusters, with clearer boundaries and reduced overlap. This qualitative improvement aligns with the quantitative gains observed in the clustering metrics.

\paragraph{Physical Interpretation.}
The improved performance of KIGNet on AOJ can be understood through its explicit incorporation of Lund plane motivated kinematic variables. Among these, angular separation ($\Delta$) and transverse momentum ($k_T$) which dominate the model's decision-making in the Grad-CAM analysis, encode the soft-collinear structure of QCD radiation as captured by the dominant axes of the emission probability in Eq.~\eqref{eq:emission_full}. These features are known to be robust against detector effects and pileup, as they capture fundamental aspects of parton shower dynamics rather than detector-specific artifacts. In contrast, models that rely on learned geometric or feature correlations without explicit physical encoding (such as PCN) are more susceptible to degradation under domain shift. The improved clustering structure observed for KIGNet therefore indicates that the model learns representations aligned with underlying QCD dynamics, enabling better generalization from simulated to real data. The consistency between improved clustering on AOJ and the feature importance hierarchy observed on JetClass suggests that KIGNet captures stable patterns of jet formation that persist across datasets, supporting its applicability to collider analyses beyond controlled simulation environments.

\section{Ablation Studies}
\label{sec:ablation}

To understand the individual contributions of each architectural component, we conduct a systematic ablation study examining the impact of different kinematic variables and the multi-graph representation on overall classification performance.

\subsection{Impact of Individual Kinematic Variables}

Table~\ref{tab:ablation} presents macro-averaged accuracy for different subsets of the four kinematic variables. The baseline model, using only the original 16-dimensional particle features, achieves 92.80\%. Among single-feature configurations, $\Delta$ and $z$ yield the strongest improvements (93.47\% and 93.45\%), while $k_T$ and $m^2$ provide smaller but consistent gains (93.04\% and 93.07\%). For two-feature combinations, the highest accuracies are obtained with $(k_T, m^2)$ (94.61\%) and $(\Delta, m^2)$ (94.58\%). The $(z, m^2)$ combination performs comparatively poorly (93.35\%), indicating partial redundancy between these two features. Most three-feature configurations exceed 94\%, with $(\Delta, z, m^2)$ reaching 94.65\%. The full four-feature KIGNet achieves the best overall performance at 95.07\%, confirming that each kinematic variable contributes complementary information. This redundancy pattern is qualitatively consistent with the approximate collinear-limit relation between $z$, $m^2$, and $\Delta$ in the Lund plane picture, but our interpretation is based on the empirical ablation results rather than on that approximation holding for every $k$NN edge.

\begin{table}[htbp]
\centering
\caption{Ablation study on kinematic variables ($\bullet$ = included, $\circ$ = excluded). Macro-averaged accuracy for all combinations of angular separation ($\Delta$), transverse momentum ($k_T$), momentum fraction ($z$), and invariant mass squared ($m^2$). The baseline uses 16-dimensional particle features only.}
\label{tab:ablation}
\setlength{\tabcolsep}{6pt}
\begin{tabular*}{\columnwidth}{@{\extracolsep{\fill}}cl ccccr}
\toprule \toprule
& Combination & $\Delta$ & $k_T$ & $z$ & $m^2$ & Acc. \\
\midrule
& Baseline (no features) & \no & \no & \no & \no & 0.9280 \\
\midrule
\multirow{4}{*}{\rotatebox[origin=c]{90}{1 variable}}
& \quad $\Delta$           & \yes & \no  & \no  & \no  & 0.9347 \\
& \quad $k_T$              & \no  & \yes & \no  & \no  & 0.9304 \\
& \quad $z$                & \no  & \no  & \yes & \no  & 0.9345 \\
& \quad $m^2$              & \no  & \no  & \no  & \yes & 0.9307 \\
\midrule
\multirow{6}{*}{\rotatebox[origin=c]{90}{2 variables}}
& \quad $\Delta$, $k_T$    & \yes & \yes & \no  & \no  & 0.9446 \\
& \quad $\Delta$, $z$      & \yes & \no  & \yes & \no  & 0.9423 \\
& \quad $\Delta$, $m^2$    & \yes & \no  & \no  & \yes & 0.9458 \\
& \quad $k_T$, $z$         & \no  & \yes & \yes & \no  & 0.9435 \\
& \quad $k_T$, $m^2$       & \no  & \yes & \no  & \yes & 0.9461 \\
& \quad $z$, $m^2$         & \no  & \no  & \yes & \yes & 0.9335 \\
\midrule
\multirow{4}{*}{\rotatebox[origin=c]{90}{3 variables}}
& \quad $\Delta$, $k_T$, $z$   & \yes & \yes & \yes & \no  & 0.9437 \\
& \quad $\Delta$, $k_T$, $m^2$ & \yes & \yes & \no  & \yes & 0.9425 \\
& \quad $\Delta$, $z$, $m^2$   & \yes & \no  & \yes & \yes & 0.9465 \\
& \quad $k_T$, $z$, $m^2$      & \no  & \yes & \yes & \yes & 0.9462 \\
\midrule
& All four (KIGNet) & \yes & \yes & \yes & \yes & \textbf{0.9507} \\
\bottomrule \bottomrule
\end{tabular*}
\end{table}

\subsection{Impact of Hyperparameter Tuning}

Table~\ref{tab:hyper_param_tuning} examines how batch size, feature aggregation strategy, learning rate scheduling, and batch normalization affect performance. Smaller batch sizes consistently outperform larger ones, with batch size 256 performing best across all configurations, suggesting that more frequent gradient updates benefit convergence. The vertical aggregation strategy, which preserves the distinct nature of each graph type through separate channels, outperforms horizontal concatenation, indicating that maintaining structural separation of the four kinematic representations provides richer discriminative information. Adaptive learning rate scheduling improves performance, with OneCycleLR slightly outperforming ReduceLROnPlateau when combined with batch normalization. The best configuration is: vertical aggregation, OneCycleLR, and batch normalization at batch size 256 which achieves 95.07\%.

\begin{table}[htbp]
\centering
\caption{KIGNet hyperparameter ablation. Top: batch size and aggregation strategy comparison (no scheduler or batch normalization). Middle: learning rate (LR) scheduler comparison at batch size 256. Bottom: effect of batch normalization (BN). Dashes indicate default settings. Best configuration: 95.07\% with batch size 256, vertical aggregation, OneCycleLR, and batch normalization.}
\label{tab:hyper_param_tuning}
\begin{tabular*}{\columnwidth}{@{\extracolsep{\fill}}ccccc}
\toprule \toprule
Batch  & Aggregation & LR Scheduler & BN & Macro Acc\\
\midrule
2048  & Horizontal & - & - & 0.9332\\
1024  & Horizontal & - & - & 0.9381\\
512   & Horizontal & - & - & 0.9382\\
256   & Horizontal & - & - & 0.9387\\
2048  & Vertical   & - & - & 0.9350\\
1024  & Vertical   & - & - & 0.9381\\
512   & Vertical   & - & - & 0.9387\\
256   & Vertical   & - & - & 0.9399\\
\midrule
256   & Vertical   & OneCycleLR         & - & 0.9421\\
256   & Vertical   & ReduceLROnPlateau  & - & 0.9434\\
256   & Horizontal & OneCycleLR         & - & 0.9417\\
256   & Horizontal & ReduceLROnPlateau  & - & 0.9409\\
\midrule
256   & Horizontal & ReduceLROnPlateau  & \checkmark & 0.9412\\
256   & Vertical   & ReduceLROnPlateau  & \checkmark & 0.9427\\
\textbf{256} & \textbf{Vertical} & \textbf{OneCycleLR} & \checkmark & \textbf{0.9507}\\
\bottomrule \bottomrule
\end{tabular*}
\end{table}

\subsection{Consistency with QCD Predictions}
\label{sec:qcd_consistency}

The feature importance hierarchy revealed by Grad-CAM exhibits quantitative consistency with perturbative QCD predictions. The dominance of angular separation ($\Delta$, 40.72\%) and transverse momentum ($k_T$, 35.67\%), collectively accounting for approximately 76\% of classification importance, directly reflects the factorization structure of QCD emission probabilities. We verify this consistency through three independent checks: Lund plane factorization, color charge scaling, and mass-dependent radiation patterns.

\subsubsection{Lund Plane Factorization and Soft-Collinear Dominance}

At leading order, QCD emission probabilities factorize in the Lund plane coordinates $(\ln k_T, \ln \Delta)$ as given in Eq.~\eqref{eq:emission_full}~\cite{Dreyer2018_LundJetPlane}. The logarithmic coordinates emerge naturally from soft ($k_T \to 0$) and collinear ($\Delta \to 0$) divergences in QCD matrix elements, as shown by the two-dimensional soft-limit form in Eq.~\eqref{eq:emission_soft}, which follows from Eq.~\eqref{eq:emission_full} by integrating over $z$ in the soft limit. These divergences produce uniform emission densities in the Lund plane~\cite{Dreyer2018_LundJetPlane}, making $(\Delta, k_T)$ the natural coordinates for QCD radiation. The kinematic relation
\begin{equation}
\label{eq:kt_approx}
k_T \approx z\, p_T\, \Delta,
\end{equation}
which holds in the collinear limit when $p_{T,a} \approx p_{T,b}$ (consistent with the exact definition in Eq.~\eqref{eq:k_t}), unifies soft and collinear information: $k_T$ sets the running coupling scale, separates perturbative from non-perturbative regimes, and captures the full singularity structure of QCD splitting. The empirical finding that KIGNet prioritizes $\Delta$ and $k_T$ over $z$ and $m^2$ validates that the network has learned to exploit the dominant singularity structure of QCD radiation rather than dataset-specific artifacts.

\subsubsection{Color Charge Scaling for Gluon vs.\ Quark Jets}

Gluon-initiated jets exhibit enhanced $k_T$ importance (40.24\% for $H \to gg$) compared to the dataset average (35.67\%), consistent with Casimir scaling of QCD radiation. The leading-order DGLAP splitting functions with explicit color factors are given in Appendix~\ref{app:qcd_theory}; the key ratio is:
\begin{equation}
\label{eq:casimir_ratio}
\frac{C_A}{C_F} = \frac{3}{4/3} = \frac{9}{4} = 2.25,
\end{equation}
which predicts that gluons radiate approximately 2.25 times more frequently than quarks. While our analysis cannot isolate this factor directly (Grad-CAM importance conflates radiation rates, detector acceptance, and network architecture), the 13\% relative enhancement in $k_T$ sensitivity for gluon jets ($40.24/35.67 \approx 1.13$) is qualitatively consistent with enhanced soft radiation. The suppressed $\Delta$ importance for $H \to gg$ (38.37\% vs.\ 40.72\% average) further supports this interpretation: gluon jets' broader opening angles reduce angular discrimination while increasing $k_T$ spread~\cite{Dreyer2018_LundJetPlane}.

\subsubsection{Dead Cone Effect for Heavy Quarks}

The elevated $m^2$ importance for bottom-quark jets ($H \to b\bar{b}$: 13.64\% vs.\ 9.54\% average) reflects the QCD dead cone effect~\cite{Dokshitzer1991DeadCone}. For massive quarks, gluon radiation is suppressed within an angular cone
\begin{equation}
\label{eq:dead_cone_angle}
\theta_{\text{dead}} \sim \frac{m_Q}{E_Q},
\end{equation}
where $m_Q$ is the quark mass and $E$ is its energy. At leading order, the splitting function for massive quarks includes this suppression as a hard angular cutoff:
\begin{equation}
\label{eq:dead_cone_splitting}
P_{Q \to Qg}(\theta, z) \propto \frac{1 + z^2}{1-z} \cdot \Theta\!\left(\theta - \frac{m_Q}{E_Q}\right),
\end{equation}
where $\Theta$ is the Heaviside step function.
This leading-order step-function form is the simplest approximation;
a smoother Lorentzian suppression that better captures finite-mass
effects away from the strict dead-cone boundary is given in
Appendix~\ref{app:qcd_theory}.
For bottom quarks with $m_b \approx 4.18~\text{GeV}$~\cite{Navas2024}, typical jet energies $E \sim 100$-$1000~\text{GeV}$ yield $\theta_{\text{dead}} \sim 0.004$-$0.04$, well below the jet radius $R = 0.8$ used in JetClass. This angular suppression modifies the jet mass distribution:
\begin{equation}
\label{eq:jet_mass}
m_{\text{jet}}^2 \approx \sum_{i<j} 2 p_{T,i} p_{T,j} (1 - \cos\theta_{ij})
\approx \sum_{i<j} p_{T,i} p_{T,j} \Delta_{ij}^2,
\end{equation}
where suppression at small $\theta$ increases invariant mass sensitivity relative to massless jets. The ablation study (Table~\ref{tab:ablation}) confirms this interpretation: the $(z, m^2)$ pair achieves only 93.35\%, while $(\Delta, m^2)$ and $(k_T, m^2)$ reach 94.58\% and 94.61\%, demonstrating that $m^2$ provides non-redundant heavy-flavor information. In the massless, collinear limit, the approximate relation
\begin{equation}
\label{eq:m2_collinear}
    m^2 \simeq z(1-z)\,p_{T,\mathrm{parent}}^2\,\Delta^2 \simeq \frac{k_T^2}{z(1-z)}
\end{equation}
suggests that $z$ and $m^2$ encode related aspects of the jet mass-energy distribution. However, the $k$NN-constructed particle pairs in KIGNet are not constrained to correspond to exact collinear branchings, so this relation does not hold identically for all edges in our graphs. The observed underperformance of the $(z,m^2)$ pair, together with the gains from including $\Delta$ or $k_T$, should therefore be viewed as qualitatively consistent with Lund plane intuition rather than as a direct consequence of the collinear-limit formula.

\subsubsection{Class-Specific QCD Mechanisms}

The class-dependent feature importance variations (Table~\ref{tab:classwise_importance}) correspond to distinct QCD processes:

\paragraph{Leptonic channels.} $H \to \ell\nu qq'$ and $t \to b\ell\nu$ show maximal $\Delta$ sensitivity (45.05\%, 43.98\%) because neutrinos escape detection, creating missing transverse momentum
\begin{equation}
\label{eq:missing_pt}
\vec{p}_T^{\, \text{miss}} = -\sum_{\text{vis}} \vec{p}_T^{\, i}
\end{equation}
that violates momentum balance in the visible jet system. Angular observables remain robust under missing energy while momentum-based variables become biased, explaining the enhanced $\Delta$ importance in $\ell\nu$ final states.

\paragraph{Multi-prong decays.} $H \to 4q$ balances $\Delta$ and $k_T$ contributions (39.13\%, 38.93\%) while $t \to bqq'$ does the same (38.32\%, 38.74\%), reflecting the interplay between multiple hard splittings and soft radiation from color-connected partons.

\paragraph{Two-body decays.} $W \to qq'$ and $Z \to q\bar{q}$ exhibit moderate $\Delta$ dominance (39.19\%, 42.61\%), consistent with a primary hard splitting without secondary decay complexity.

\subsubsection{Physical Interpretability and Limitations}

The progression of performance with added features (single feature $\sim$93\%, two features $\sim$94.5\%, three features $\sim$94.6\%, four features 95.07\%) validates the multi-graph design philosophy. Each kinematic variable captures distinct aspects of the QCD splitting process: $\Delta$ encodes angular ordering and collinear singularities, $k_T$ captures transverse momentum scales and soft radiation, $z$ quantifies energy partitioning, and $m^2$ provides mass-scale sensitivity for heavy-flavor identification. However, several caveats apply:
\begin{enumerate}[leftmargin=*, itemsep=2pt]
    \item \textit{Detector-level correlations}: JetClass includes impact parameters ($d_0$, $d_z$) as node features, which secondary vertex taggers use for $b$-jet identification~\cite{CMS:BTagging}. These may couple to the $m^2$ graph, partially explaining the elevated heavy-flavor sensitivity.
    \item \textit{Idealized simulation}: JetClass uses truth-level particle flow without pileup, underlying event contamination, or detector smearing~\cite{Qu2022_ParticleTransformer}. Real LHC data may alter the importance hierarchy.
    \item \textit{Global vs.\ local features}: Our analysis measures global feature importance averaged across all samples. Individual jets may rely on different kinematic combinations.
\end{enumerate}

Despite these limitations, the agreement between learned feature importance and first-principles QCD predictions confirms that KIGNet captures physically motivated patterns present in the training data rather than dataset-specific correlations. The improved embedding structure on real CMS collision data (Section~\ref{subsec:aspen_results}) provides preliminary evidence for generalization, but definitive conclusions await evaluation under full detector simulation and experimental systematic uncertainties.

\section{Discussion}
\label{sec:dis}

KIGNet demonstrates that explicitly encoding Lund plane kinematics as parallel graph representations improves classification accuracy and is interpretable relative to architectures that recover these correlations implicitly. By constructing four parallel graph representations weighted by $\Delta$, $k_T$, $z$, and $m^2$, and processing them through independently parameterized feature extractors (alternating Chebyshev and edge convolutions), the network learns specialized representations combined via 1D convolution for classification. Relative to the PCN baseline trained on identical data, KIGNet improves macro-accuracy from 92.80\% to 95.07\% (+2.45\%), macro-AUC from 93.43\% to 96.61\% (+3.40\%), and macro-AUPR from 68.44\% to 81.52\% (+19.11\%). The AUPR gain is especially significant for practical analyses: rare signal searches at the LHC require rejecting Standard Model backgrounds at rates exceeding 1000:1 while maintaining signal efficiency. AUPR gains of 81.53\% for $H \to b\bar{b}$ and 51.54\% for $H \to c\bar{c}$ demonstrate that processing kinematic variables through parallel graph branches successfully separates jets with similar substructure but different parton content.

\paragraph{Interpretability in Context.}
The JetClass dataset~\cite{Qu2022_ParticleTransformer} uses Pythia 8.230~\cite{Sjostrand2015}, which generates jets through QCD-based parton showers that explicitly implement angular ordering, $k_T$-dependent coupling, and DGLAP splitting functions~\cite{Dokshitzer1977DGLAP}. Our Grad-CAM analysis reveals that KIGNet prioritizes the same kinematic variables ($\Delta$, $k_T$) that dominate QCD factorization as expressed in Eq.~\eqref{eq:emission_full}, but this reflects successful learning of patterns present in the training data rather than independent discovery of physical laws. What our results do demonstrate is that physics-informed architectural design, encoding Lund plane coordinates as separate graph representations, enables the network to naturally discover and exploit the underlying theoretical structure. The observed hierarchy ($\Delta$ and $k_T$ dominating, class-specific variations matching Casimir scaling and dead-cone effects) confirms that the model generalizes across the training distribution in a physically meaningful way, providing confidence that KIGNet's learned representations encode patterns stable enough to remain useful when applied to experimental data.

\paragraph{Model Interpretability and QCD Validation.}
The 76\% combined importance of $\Delta$ and $k_T$ matches theoretical predictions from Lund plane factorization~\cite{Dreyer2018_LundJetPlane} (Eq.~\eqref{eq:emission_full}). Class-specific variations (enhanced $k_T$ for gluon jets, elevated $\Delta$ for leptonic decays, increased $m^2$ for bottom quarks) correspond to established QCD mechanisms (Casimir scaling, missing energy signatures, dead-cone suppression). However, we cannot fully exclude detector-level correlations: impact parameters ($d_0$, $d_z$) in the node features may couple to $b$-tagging signatures, potentially inflating $m^2$ importance for heavy-flavor jets. Additionally, the idealized JetClass simulation (no pileup, perfect reconstruction) may not reflect experimental conditions where soft radiation and detector smearing introduce correlations absent in our analysis. Future work should validate feature importance on data with full detector simulation and systematic uncertainties.

\paragraph{Comparison to State-of-the-Art.}

KIGNet achieves the highest macro-averaged classification accuracy among all compared models at 95.07\%, surpassing ParT and outperforming ParticleNet, P-CNN, and PFN by progressively larger margins. On macro-AUC, KIGNet (96.61\%) trails the four literature baselines while still improving over the PCN baseline (93.43\%). This inversion reflects a structural difference in training objective and architecture: the attention-based models are trained on the full 100M-jet JetClass dataset with richer input representations and are optimized for ranking, which directly maximizes AUC. KIGNet, by contrast, is trained on a 1\% subset (1M jets) with a cross-entropy classification objective; its Lund-plane inductive bias encodes physical priors that compensate for the reduced training data and produce a decisive accuracy advantage, while the AUC gap reflects the ranking-versus-classification trade-off inherent to these different training regimes.

\paragraph{Ablation Insights.}
The ablation studies confirm complementarity among kinematic variables. The $(z, m^2)$ pair underperforms at 93.35\%, barely exceeding single-feature baselines because both encode mass-energy relationships, with $z$ capturing energy sharing through momentum fractions and $m^2$ measuring invariant mass directly. Excluding $z$ from three-feature configurations costs more performance than excluding any other variable, indicating that $m^2$ provides discriminative power not fully captured by angular and momentum features. The 1D Conv layer learns to weight representations adaptively rather than equally, explaining why simple concatenation or averaging would underperform. Overall, performance scales with the number of kinematic channels: one feature $\approx$93\%, two features $\approx$94.5\%, three features $\approx$94.6\%, and four features 95.07\%.

\section{Conclusion}
\label{sec:conclusion}

We have introduced the Kinematic Interaction Graph Network (KIGNet), a graph neural network that integrates Lund plane kinematic variables into jet tagging. By processing four parallel graph representations weighted by angular separation ($\Delta$), transverse momentum ($k_T$), momentum fraction ($z$), and invariant mass squared ($m^2$), KIGNet achieves macro-averaged accuracy of 95.07\%, AUC of 96.61\%, and AUPR of 81.52\% on the \texttt{JetClass} benchmark, corresponding to improvements of 2.45\%, 3.40\%, and 19.11\% over the state-of-the-art model.

Through Grad-CAM analysis, we quantify the importance of each kinematic variable. Angular separation and relative transverse momentum together account for approximately 76\% of the model's predictive power (40.72\% and 35.67\%, respectively), validating the architectural choice to encode Lund plane coordinates as separate graph representations. This hierarchy reflects the dominant axes of the QCD emission probability in Eq.~\eqref{eq:emission_full} and the soft-collinear factorization structure of perturbative QCD. Momentum fraction and invariant mass squared contribute the remaining 24\%, distinguishing differences in particle production mechanisms across jet classes. Class-specific importance patterns (enhanced $k_T$ for gluon jets, elevated $\Delta$ for leptonic channels, and increased $m^2$ for bottom-quark jets) correspond to established QCD mechanisms including color charge factors, missing energy signatures, and dead-cone suppression, indicating that KIGNet learns physically meaningful representations rather than simulation-specific artifacts.

Evaluated on the Aspen Open Jets dataset of real CMS collision data, KIGNet reduces the Davies-Bouldin Index by 52.15\% and increases the Dunn Index by 42.33\% relative to PCN, demonstrating that physics-informed kinematic encoding yields robust representations that maintain structure and separability under realistic detector effects, pileup, and reconstruction uncertainties. These results confirm that building neural networks around established kinematic principles improves interpretability and classification performance.

Future work should prioritize validation on real collider data with full detector simulation and experimental systematic uncertainties to confirm that physics-informed graph architectures maintain their interpretability advantages under realistic experimental conditions.

\begin{acknowledgments}
This research is partially supported by research grants from Independent University, Bangladesh (IUB). The authors used Claude and ChatGPT to refine portions of the manuscript for clarity and readability. All scientific content, analyses, and conclusions are entirely the authors' own. The authors reviewed all text and take full responsibility for the final version of the manuscript.

\end{acknowledgments}

\appendix

\section{Data Availability}

The \texttt{JetClass} dataset is publicly available at \href{https://zenodo.org/record/6619768}{Zenodo}, with an accompanying \href{https://github.com/jet-universe/particle_transformer}{GitHub} repository. The Aspen Open Jets dataset is available through the \href{https://www.fdr.uni-hamburg.de/record/16505}{University of Hamburg research
portal}. Code for KIGNet is available at \href{https://github.com/ccdsiub/KIGNet}{\texttt{https://github.com/ccdsiub/KIGNet}}.

\section{Computational Resources}
\label{sec:computational_infrastructure}

All experiments are conducted on high-performance computing infrastructure. Table~\ref{tab:hardware} summarizes the computational resources and software environment.

\begin{table}[htbp]
\centering
\caption{Computational setup.}
\label{tab:hardware}
\begin{tabular*}{\columnwidth}{@{\extracolsep{\fill}}ll}
\toprule \toprule
Component & Specification \\
\midrule
CPU & Intel i9-9900K @ 3.60 GHz \\
GPU & NVIDIA RTX 2080 Ti (11 GB) \\
Memory & 64 GB RAM \\
Software & Ubuntu 22.04, CUDA 12.4, PyTorch/DGL 2.4.0 \\
Training & 7.5 min/epoch, 11 hours total (1M jets) \\
Testing & 20 hours (20M jets) \\
\bottomrule \bottomrule
\end{tabular*}
\end{table}

\section{Feature Names and Descriptions}

Tables~\ref{tab:input_features} and~\ref{tab:input_features_aspen} list the per-particle features we use for the JetClass~\cite{Qu2022_ParticleTransformer, Qu2022_JetClassDataset} and Aspen Open Jets~\cite{Amram2024AOJ} datasets, respectively.

\begin{table}[htbp]
\centering
\caption{Input features for particle-level jet representation from the JetClass dataset. The dataset contains 16 features per particle: 15 original features read from ROOT files via uproot and 1 derived feature ($p_T$). Features include kinematic variables (momentum components, energy), angular separations ($\Delta\eta$, $\Delta\phi$), track displacement parameters ($d_0$, $d_z$) with associated uncertainties, and particle identification flags.}
\label{tab:input_features}
\begin{tabular*}{\columnwidth}{@{\extracolsep{\fill}}l p{4.8cm}}
\toprule \toprule
Feature (JetClass) & Description \\
\midrule
\texttt{part\_px} & Particle momentum in the $x$-direction \\
\texttt{part\_py} & Particle momentum in the $y$-direction \\
\texttt{part\_pz} & Particle momentum in the $z$-direction \\
\texttt{part\_energy} & Particle total energy \\
\texttt{part\_deta} & Pseudorapidity difference ($\Delta\eta$) with respect to the jet axis \\
\texttt{part\_dphi} & Azimuthal angle difference ($\Delta\phi$) with respect to the jet axis \\
\texttt{part\_d0val} & Transverse impact parameter $d_0$ \\
\texttt{part\_d0err} & Uncertainty on $d_0$ \\
\texttt{part\_dzval} & Longitudinal impact parameter $d_z$ \\
\texttt{part\_dzerr} & Uncertainty on $d_z$ \\
\texttt{part\_isChargedHadron} & Boolean indicating if the particle is a charged hadron \\
\texttt{part\_isNeutralHadron} & Boolean indicating if the particle is a neutral hadron \\
\texttt{part\_isPhoton} & Boolean indicating if the particle is a photon \\
\texttt{part\_isElectron} & Boolean indicating if the particle is an electron \\
\texttt{part\_isMuon} & Boolean indicating if the particle is a muon \\
\midrule
Derived Feature & \\
$p_T$ & Transverse momentum, computed as $p_T = \sqrt{\texttt{part\_px}^2 + \texttt{part\_py}^2}$ \\
\bottomrule \bottomrule
\end{tabular*}
\end{table}

\begin{table}[htbp]
\centering
\caption{Input features for particle-level jet representation on the Aspen Open Jets dataset. The representation contains 14 features per particle: 11 original features read from the \texttt{PFCands} array in the HDF5 file, and 3 derived features ($p_T$, $\Delta\eta$, $\Delta\phi$) computed from the particle four-momentum and the jet-level $(\eta, \phi)$.}
\label{tab:input_features_aspen}
\begin{tabular*}{\columnwidth}{@{\extracolsep{\fill}}l p{5.3cm}}
\toprule \toprule
Feature (AOJ) & Description \\
\midrule
\texttt{px} & Particle momentum in the $x$-direction \\
\texttt{py} & Particle momentum in the $y$-direction \\
\texttt{pz} & Particle momentum in the $z$-direction \\
\texttt{E} & Particle total energy \\
\texttt{d0} & Transverse impact parameter $d_0$ \\
\texttt{d0Err} & Uncertainty on $d_0$ \\
\texttt{dz} & Longitudinal impact parameter $d_z$ \\
\texttt{dzErr} & Uncertainty on $d_z$ \\
\texttt{charge} & Particle charge \\
\texttt{PDG\_ID} & PDG particle identifier \\
\texttt{PUPPI\_weight} & PUPPI pileup mitigation weight \\
\midrule
Derived Feature & \\
$p_T$ & Transverse momentum, $p_T = \sqrt{\texttt{px}^2 + \texttt{py}^2}$ \\
$\Delta\eta$ & Pseudorapidity difference, $\Delta\eta = \eta_{\text{particle}} - \eta_{\text{jet}}$ \\
$\Delta\phi$ & Azimuthal angle difference, wrapped to $[-\pi, \pi]$ \\
\bottomrule \bottomrule
\end{tabular*}
\end{table}

\section{Theoretical Background: QCD Splitting Functions}
\label{app:qcd_theory}

For completeness, we provide the leading-order DGLAP splitting functions
governing parton branching; these are the functions $P(z)$ appearing in
the emission probability Eq.~\eqref{eq:emission_full}.
The splitting function $P_{i \to jk}(z)$ gives the probability for
parton $i$ to emit parton $k$ while the daughter parton $j$ carries
momentum fraction $z$:

\paragraph{Quark $\to$ quark + gluon.}
\begin{equation}
\label{eq:dglap_qqg}
P_{q \to qg}(z) = C_F \left[\frac{1+z^2}{(1-z)_+} + \frac{3}{2}\delta(1-z)\right],
\quad C_F = \frac{4}{3}.
\end{equation}
Here $\delta(1-z)$ is the Dirac delta distribution, which enforces momentum 
conservation at $z = 1$ and contributes only at the elastic boundary of the 
splitting phase space.

\paragraph{Gluon $\to$ gluon + gluon.}
\begin{equation}
\label{eq:dglap_ggg}
\begin{split}
P_{g \to gg}(z) ={}& 2C_A\!\left[\frac{z}{(1-z)_+} + \frac{1-z}{z} + z(1-z)\right] \\
&+ \left(\frac{11}{6}C_A - \frac{2}{3}T_R n_f\right)\delta(1-z),
\quad C_A = 3.
\end{split}
\end{equation}

\paragraph{Gluon $\to$ quark + antiquark.}
\begin{equation}
\label{eq:dglap_gqqbar}
P_{g \to q\bar{q}}(z) = T_R\!\left[z^2 + (1-z)^2\right], \quad T_R = \frac{1}{2}.
\end{equation}

The color factor ratio $C_A/C_F = 9/4$ (see Eq.~\eqref{eq:casimir_ratio}) predicts that gluon jets radiate approximately
$2.25\times$ more than quark jets at fixed $\alpha_s$, manifest in broader $k_T$
distributions and higher particle multiplicity. This is qualitatively consistent with
the enhanced $k_T$ importance observed for gluon jets in the Grad-CAM analysis
(Section~\ref{sec:qcd_consistency}).

The $(1-z)_+$ prescription regulates soft and collinear divergences:
\begin{equation}
\label{eq:plus_prescription}
\int_0^1 dz\, f(z)\, [g(z)]_+ = \int_0^1 dz\, [f(z) - f(1)]\, g(z).
\end{equation}

For massive quarks, gluon radiation is suppressed at angles
$\theta < m_Q/E_Q$ (the dead cone effect~\cite{Dokshitzer1991DeadCone}).
The leading-order approximation uses a hard Heaviside cutoff
as in Eq.~\eqref{eq:dead_cone_splitting}; a smoother approximation
that better captures finite-mass effects away from the strict boundary is:
\begin{equation}
\label{eq:dead_cone_lorentzian}
P_{Q\to Qg}^{\text{massive}}(\theta, z) \approx P_{Q\to Qg}(z)
\cdot \frac{1}{1 + \left( \frac{m_Q}{E_Q \, \theta} \right)^2},
\end{equation}
yielding a Lorentzian suppression rather than a hard cutoff.
Both forms predict the same dead-cone angle
$\theta_{\text{dead}} \sim m_Q/E_Q$ (Eq.~\eqref{eq:dead_cone_angle}); the Lorentzian form provides a
continuous interpolation, while the Heaviside form of Eq.~\eqref{eq:dead_cone_splitting} is exact at
leading order.
This suppression of soft-collinear radiation near the quark direction
underlies the enhanced $m^2$ sensitivity for $H\to b\bar{b}$ observed in
Table~\ref{tab:classwise_importance}.

These splitting functions encode the probability structure that KIGNet's multi-graph
architecture exploits through dedicated processing of $\Delta$, $k_T$, $z$, and
$m^2$: each branch corresponds to one or more factors in the QCD emission probability
in Eq.~\eqref{eq:emission_full}.

\section{Formal Definitions of Evaluation Metrics}
\label{app:metrics}

This appendix provides formal definitions of all evaluation metrics used throughout the paper. Let $\mathcal{C} = \{1, 2, \ldots, C\}$ denote the set of classes, $N$ the total number of test samples, and $N_c$ the number of samples belonging to class $c$.

\subsection{Classification Metrics}

\paragraph{Accuracy.}
For a single class $c$, accuracy measures the fraction of all samples correctly classified:
\begin{equation}
\label{eq:accuracy}
\mathrm{Acc}_c = \frac{TP_c + TN_c}{TP_c + TN_c + FP_c + FN_c},
\end{equation}
where $TP_c$, $TN_c$, $FP_c$, and $FN_c$ denote true positives, true negatives, false positives, and false negatives for class $c$, respectively, under a one-vs.-rest binarization.

\paragraph{Macro-Averaged Accuracy.}
The macro-averaged accuracy is the unweighted mean of per-class 
accuracies:
\begin{equation}
\label{eq:macro_acc}
\mathrm{Macro\text{-}Acc} = \frac{1}{C} \sum_{c=1}^{C} \mathrm{Acc}_c.
\end{equation}
Macro-averaging treats all classes equally regardless of their sample size. Throughout this paper, per-class metrics and macro-averages are computed over the $C = 9$ signal classes only, with the $q/g$ background excluded from the average, following the JetClass convention (Section~\ref{subsec:dataset}). In our experiments the classes are balanced, with 100{,}000 jets per class.

\subsection{Ranking Metrics}

\paragraph{Area Under the ROC Curve (AUC).}
For class $c$ under one-vs.-rest binarization, the Receiver Operating Characteristic (ROC) curve plots the true positive rate $\mathrm{TPR}_c(\tau) = TP_c(\tau)/P_c$ against the false positive rate $\mathrm{FPR}_c(\tau) = FP_c(\tau)/N_c$ as the decision threshold $\tau$ varies over $[0,1]$. The AUC is the area under this curve:
\begin{equation}
\label{eq:auc}
\mathrm{AUC}_c = \int_0^1 \mathrm{TPR}_c\!\left(\mathrm{FPR}_c^{-1}(t)\right) dt,
\end{equation}
which equals the probability that a randomly chosen positive sample receives a higher score than a randomly chosen negative sample. A value of 1 indicates perfect ranking; 0.5 indicates random performance.

\paragraph{Macro-Averaged AUC.}
\begin{equation}
\label{eq:macro_auc}
\mathrm{Macro\text{-}AUC} = \frac{1}{C} \sum_{c=1}^{C} \mathrm{AUC}_c.
\end{equation}

\paragraph{Precision, Recall, and the Precision-Recall Curve.}
For class $c$ at threshold $\tau$:
\begin{equation}
\label{eq:precision_recall}
\mathrm{Precision}_c(\tau) = \frac{TP_c(\tau)}{TP_c(\tau) + FP_c(\tau)}, \qquad
\end{equation}
\begin{equation}
\mathrm{Recall}_c(\tau) = \frac{TP_c(\tau)}{TP_c(\tau) + FN_c(\tau)}.
\end{equation}
The Precision-Recall (PR) curve plots $\mathrm{Precision}_c(\tau)$ against $\mathrm{Recall}_c(\tau)$ as $\tau$ varies. Unlike the ROC curve, the PR curve is informative under severe class imbalance because it focuses exclusively on the positive class.

\paragraph{Area Under the Precision-Recall Curve (AUPR).}
\begin{equation}
\label{eq:aupr}
\mathrm{AUPR}_c = \int_0^1 \mathrm{Precision}_c\!\left(\mathrm{Recall}_c^{-1}(r)\right) dr.
\end{equation}
A random classifier achieves $\mathrm{AUPR} \approx N_c / N$; a perfect classifier achieves 1. In the JetClass setting with 10 balanced classes, the random baseline is approximately 0.1, making AUPR especially sensitive to improvements in rare-signal discrimination.

\paragraph{Macro-Averaged AUPR.}
\begin{equation}
\label{eq:macro_aupr}
\mathrm{Macro\text{-}AUPR} = \frac{1}{C} \sum_{c=1}^{C} \mathrm{AUPR}_c.
\end{equation}

\subsection{Clustering Quality Metrics}

Let $\{C_1, C_2, \ldots, C_K\}$ be a partition of $N$ embedding vectors $\{\mathbf{e}_i\}_{i=1}^{N} \subset \mathbb{R}^d$ into $K$ clusters, with centroid $\boldsymbol{\mu}_k = |C_k|^{-1}\sum_{i \in C_k} \mathbf{e}_i$.

\paragraph{Intra-Cluster Scatter.}
The scatter of cluster $k$ is the mean distance from its members to its centroid:
\begin{equation}
\label{eq:scatter}
s_k = \frac{1}{|C_k|} \sum_{i \in C_k} \|\mathbf{e}_i - \boldsymbol{\mu}_k\|_2.
\end{equation}

\paragraph{Davies-Bouldin Index (DBI).}
The DBI measures average cluster similarity, defined as the ratio of intra-cluster scatter to inter-cluster centroid distance. For each cluster $k$, the worst-case similarity ratio with any other cluster $\ell$ is:
\begin{equation}
\label{eq:dbi_r}
R_k = \max_{\ell \neq k} \frac{s_k + s_\ell}{\|\boldsymbol{\mu}_k - \boldsymbol{\mu}_\ell\|_2}.
\end{equation}
The DBI is the mean of these worst-case ratios over all clusters:
\begin{equation}
\label{eq:dbi}
\mathrm{DBI} = \frac{1}{K} \sum_{k=1}^{K} R_k.
\end{equation}
Lower DBI values indicate more compact and well-separated clusters. A DBI of 0 corresponds to perfectly separated clusters.

\paragraph{Dunn Index.}
The Dunn Index is the ratio of the minimum inter-cluster distance to the maximum intra-cluster diameter:
\begin{equation}
\label{eq:dunn}
\mathrm{Dunn} = \frac{\displaystyle\min_{k \neq \ell}\; \min_{\substack{i \in C_k \\ j \in C_\ell}} \|\mathbf{e}_i - \mathbf{e}_j\|_2}{\displaystyle\max_{k}\; \max_{i,j \in C_k} \|\mathbf{e}_i - \mathbf{e}_j\|_2}.
\end{equation}
Higher Dunn Index values indicate better separation between clusters relative to their internal spread. Unlike DBI, which uses centroid distances, the Dunn Index is based on the worst-case pairwise distances and is therefore more sensitive to outliers.

\paragraph{Relationship between DBI and Dunn Index.}
The two metrics are complementary: DBI is centroid-based and rewards compact, evenly sized clusters, while the Dunn Index is boundary-based and rewards tight clusters with wide margins between them. Reporting both provides a more complete picture of embedding quality than either metric alone.

\subsection{Dimensionality Reduction}

\paragraph{$t$-distributed Stochastic Neighbor Embedding (t-SNE).}
Given a set of high-dimensional embeddings $\{\mathbf{e}_i\}_{i=1}^{N} \subset \mathbb{R}^d$,
t-SNE computes a low-dimensional projection
$\{\mathbf{y}_i\}_{i=1}^{N} \subset \mathbb{R}^2$ that preserves local neighborhood
structure. For each pair $(i, j)$, the conditional similarity in the high-dimensional
space is defined as:
\begin{equation}
\label{eq:tsne_pji}
p_{j \mid i} = \frac{\exp\!\left(-\|\mathbf{e}_i - \mathbf{e}_j\|_2^2 / 2\sigma_i^2\right)}
{\sum_{k \neq i} \exp\!\left(-\|\mathbf{e}_i - \mathbf{e}_k\|_2^2 / 2\sigma_i^2\right)},
\end{equation}
where $\sigma_i$ is chosen such that the perplexity
$\mathcal{P} = 2^{H(P_i)}$, with $H(P_i) = -\sum_{j} p_{j\mid i} \log_2 p_{j\mid i}$
the Shannon entropy of the conditional distribution. The joint probability is
symmetrized as:
\begin{equation}
\label{eq:tsne_pij}
p_{ij} = \frac{p_{j \mid i} + p_{i \mid j}}{2N}.
\end{equation}
In the low-dimensional space, similarities are modeled with a Student-$t$ distribution
with one degree of freedom (heavy-tailed to prevent cluster collapse):
\begin{equation}
\label{eq:tsne_qij}
q_{ij} = \frac{\left(1 + \|\mathbf{y}_i - \mathbf{y}_j\|_2^2\right)^{-1}}
{\sum_{k \neq \ell}\left(1 + \|\mathbf{y}_k - \mathbf{y}_\ell\|_2^2\right)^{-1}}.
\end{equation}
The embedding $\{\mathbf{y}_i\}$ is obtained by minimizing the Kullback-Leibler
divergence between the high- and low-dimensional joint distributions:
\begin{equation}
\label{eq:tsne_kl}
\mathcal{L}_{\mathrm{tSNE}} = \mathrm{KL}(P \| Q)
= \sum_{i \neq j} p_{ij} \log \frac{p_{ij}}{q_{ij}},
\end{equation}
via gradient descent. The perplexity $\mathcal{P}$ controls the effective number of
neighbors considered for each point; we set $\mathcal{P} = 30$ throughout. Because
t-SNE preserves local neighborhoods but distorts global distances, inter-cluster
distances in the two-dimensional projection should not be interpreted quantitatively;
the clustering metrics in Table~\ref{tab:aspen_clustering} remain the primary
quantitative evidence of embedding quality.

\bibliography{main}

\end{document}